\documentclass[12pt,12pt]{article}
\usepackage{graphicx}
\usepackage{epsfig}
\usepackage{amsmath,amssymb}

\catcode`\@=11
\newdimen\ex@
\font\dozeb=cmmib10 scaled \magstep1
\font\dozesyb=cmbsy10 scaled \magstep1
\font\dezb=cmmib10
\def\bm{\fam9}
\def\beq{\begin{equation}}
\def\eeq{\end{equation}}
\def\beqa{\begin{eqnarray}}
\def\eeqa{\end{eqnarray}}
\newcommand\BA{\begin{array}}
\newcommand\EA{\end{array}}

\begin{document}

\title{ \bf Description of  Collective Motion in\\
\vspace{0.3cm}
Two-Dimensional Nuclei; \\
\vspace{0.3cm}
Tomonaga's Method Revisited}
\author{Seiya NISHIYAMA\footnotemark[1]~
and Jo\~{a}o da PROVID\^{E}NCIA\footnotemark[2]\\
\\
Centro de F\'\i sica,
Departamento de F\'\i sica,\\
\\
Universidade de Coimbra,
P-3004-516 Coimbra, Portugal\\[0.5cm]
{\it Dedicated to the Memory of Toshio Marumori}}

\def\bm#1{\mbox{\boldmath $#1$}}
\def\bra#1{\langle #1 |}
\def\ket#1{| #1 \rangle}

\maketitle
\vspace{15mm}
\footnotetext[1]
{Corresponding author. 
~E-mail address: 
seikoceu@khe.biglobe.ne.jp,~
nisiyama@teor.fis.uc.pt}
\footnotetext[2]
{E-mail address: providencia@teor.fis.uc.pt}

\begin{abstract}
Four decades ago, Tomonaga proposed 
the elementary theory of quantum mechanical 
collective motion of two-dimensional nuclei of 
$N$ nucleons.
The theory is based essentially on neglecting
$\frac{1}{\sqrt N}$ against unity.
Very recently we have given
$exact$ canonically conjugate momenta
to quadrupole-type collective coordinates
under some subsidiary conditions and
have derived nuclear quadrupole-type collective Hamiltonian.
Even in the case of simple two-dimensional nuclei,
we require a subsidiary condition to obtain $exact$ canonical variables.
Particularly the structure of the collective subspace
satisfying the subsidiary condition is studied in detail.
This subsidiary condition
is important to investigate the structure of
the collective subspace.
\end{abstract}

\vspace{0.1cm}

{\it Keywords}:

Collective motion of two-dimensional nuclei;$\!$
Exact canonically conjugate momenta

\newpage


\def\thesection{\arabic{section}}

\section{Introduction}

\vspace{-0.5cm}

~~~In studies of collective motions in nuclei, the very difficult 
problems of large-amplitude collective motions, 
which are strongly non-linear 
phenomena in quantum nuclear dynamics, still remain unsolved. 
How do we go beyond the usual mean field theories towards 
the construction of a theory for 
large-amplitude collective motions in nuclei?
A proper treatment of collective variable was attempted.
Four decades ago, Tomonaga proposed 
the elementary theory of quantum mechanical 
collective motion of two-dimensional nuclei of 
$N$ nucleons.
The theory is based essentially on the 
neglect of 
$\frac{1}{\sqrt N}$ against unity
\cite{Tomo.55}.
Marumori $et~al$. first gave a foundation of the unified model
of collective motion and independent particle motion in nuclei
and further investigated the collective motion from the standpoint
of particle excitations
\cite{Maru.60}.

Applying Tomonaga's basic idea in his collective motion theory 
to nuclei with the aid of the Sunakawa's integral equation method
\cite{SYN.62},
one of the present authors (S.N.) 
developed a collective description of surface oscillations of nuclei
\cite{Nishi.77}.
This description is considered to give a possible microscopic 
foundation of nuclear collective motion derived from the famous 
Bohr-Mottelson model
\cite{BM.74}
(see textbooks
\cite{EG.87,RoweWood.2010}).
Introducing appropriate collective 
variables, this collective description was formulated by using the first 
quantized language, contrary to the second quantized manner
in the Sunakawa method.
Preceding the previous work
\cite{Nishi.77},
extending the Tomonaga's idea 
to three-dimensional case,
Miyazima-Tamura
\cite{MiyaTamu.56,Tamura.56}
successfully proposed a collective 
description of the surface oscillations of nuclei.
As they already pointed out, however,
there exist two serious difficulties still remaining 
in traditional theoretical treatments of the nuclear collective motions: 
(i)   Collective momenta defined according Tomonaga's approach are not exact 
canonically conjugate to collective coordinates;  
(ii)  The collective momenta are not independent from each other.   

Very recently we have given
$exact$ canonically conjugate momenta
to
quadrupole-type collective coordinates and
have derived a nuclear quadrupole-type collective Hamiltonian
\cite{NPA.14}
(referred to as I). 
The $exact$ canonically conjugate momenta
$\Pi_{2 \mu}$
to the quadrupole-type collective coordinates
$\phi_{2 \mu}$
is derived by modifying the approximate momenta
$\pi_{2 \mu}$
adopted by Miyazima-Tamura
\cite{MiyaTamu.56}
with the use of the discrete version of the Sunakawa's integral equation
\cite{SYN.62}.
We have shown that
the $exact$ canonical commutation relations between
the collective variables
$\phi_{2 \mu}$
and
$\Pi_{2 \mu}$
and the commutativity of the momenta
$\Pi_{2 \mu}$ and $\Pi_{2 \mu'}$
under some subsidiary conditions
are satisfied.
Using the $exact$ canonical variables
$\phi_{2 \mu}$
and
$\Pi_{2 \mu}$,
we found the collective Hamiltonian
which includes the so-called surface phonon-phonon interaction
corresponding to the Hamiltonian of Bohr-Mottelson model.
Even in the case of simple two-dimensional nuclei,
we have a subsidiary condition to obtain $exact$ canonical variables.
Particularly,
we study the structure of the collective subspace
satisfying the subsidiary condition.
This condition
is important to investigate the structure of
the collective subspace.
This is an interesting problem
which is solved easily 
due to the simplicity of the two-dimensional nuclei.

In Section 2,
we introduce collective variables $\phi_i$
in two-dimensional nuclei
and
approximate conjugate momenta $\eta_i$.
In Section 3,
we define the $exact$ canonically conjugate momenta $\pi_i$
and prove the commutativity of the $exact$ collective momenta 
on the collective subspace.
In Section 4,
the
$exact$ $\pi_i$-dependent kinetic part $T$ of Hamiltonian is derived.
The $\phi_i$-dependent kinetic part $T$ of the Hamiltonian
including  the constant term
is also given.
In the last section, we discuss the subsidiary condition
and give some concluding remarks.
In Appendix A, 
we derive some approximate relations for the collective conjugate momenta
which play crucial roles to determine the $\phi_i$-dependence of the $T$.

\newpage


\def\thesection{\arabic{section}}
\setcounter{equation}{0}
\renewcommand{\theequation}{\arabic{section}.
\arabic{equation}}     

\section{Collective variables and associated relations}

~~~~~For the sake of simplicity, we focus on collective motions in the 
two-dimensional nuclei. The present illustration is oversimplified 
as far as we consider only the two-dimensional case.

Consider a two-dimensional nucleus consisting of $N$ nucleons 
interacting strongly with each other. Let us denote the coordinates 
of $n$-th nucleon by $( x_n, y_n )$ and by $(p_{xn}, p_{yn})$ 
their conjugate momenta, respectively.

The total  Hamiltonian $H$ of our system is given by\\[-10pt]
\beqa
\BA{rl}
H = T + V 
= - {\displaystyle \frac{\hbar ^2}{2 \mu}}
\sum_{n=1}^N \!
\left( 
{\displaystyle
\frac{\partial ^2}{\partial x_n ^2} 
+
\frac{\partial ^2}{\partial y_n ^2}
}
\right) 
+ V(x_1, y_1; x_2, y_2;\cdots ; x_N, y_N )~,
\EA
\label{Hamiltonian}
\eeqa
where
$
T
=
1/2m
\sum_{n=1}^N (p_{xn}^2 + p_{yn}^2 )
$ 
is the total kinetic energy and $V$ is the interacting potential 
depending only on relative coordinates of the nucleons.

Following Miyazima-Tamura
\cite{MiyaTamu.56},
we introduce
two collective coordinates\\[-12pt] 
\beqa
\BA{c}
\phi_1 = {\displaystyle \frac{1}{N r_0 ^2}} 
\sum_{n=1}^N
{\displaystyle \frac{1}{2}}
(x_n ^2 - y_n ^2 ) ,~~~
\phi_2 = {\displaystyle \frac{1}{N r_0 ^2}} 
\sum_{n=1}^N x_n y_n ~,
\EA
\label{Collectivecoordinates}
\eeqa
where $r_0$ is defined as
$r_0 ^2 \!=\! \frac{R_0 ^2}{2}$ 
and $R_0$ means the nuclear equilibrium radius.
Those collective coordinates correspond to velocity potentials
from which a surface collective-motion is derived.
This is a low energy excitation most likely
occurring in the two-dimensional nuclei.
The collective variable $\phi_1$ characterizes an ellipsoidal deformation
with axis along the coordinate axis and $\phi_2$ with axis at $45^{\circ}$.
Their geometric interpretation is that
the convenient canonically
conjugate coordinate to the total angular momentum operator $l$,
i.e., the second equation of
(\ref{r**2andAngularmomentum})
is connected with the orientation of the axis of inertia and is given by
$\frac{1}{2}\arctan \frac{\phi_2}{\phi_1}$.
This was already pointed out by Tamura
\cite{Tamura.56}.
In a mean field description of the system described the the present model,
two situations may be met, either at the minimum energy the expectation
values of the operators $\phi_1$ and $\phi_2$ vanish or not. In the first
case a vibrational regime prevails while in the second case it is the
rotational regime which is favored. In the present model, it is expected
that the vibrational regime occurs for lower mass nuclei, while the
rotational regime is expected to arise for higher mass systems, the
tendency for deformation and occurrence of a rotational regime
increasing with the size. The general feature of the collective potential
energy will be very much like a Mexican hat.
 
Following Tomonaga
\cite{Tomo.55},
we introduce  
collective conjugate momenta to $\phi_1$ and $\phi_2$ 
in the sense of Tomonaga through\\[-20pt]
\beqa
\eta _i = \mu N r_0 ^2 \dot{\phi}_i
= \mu N r_0 ^2 \frac{i}{\hbar} [T, \phi _i ]~, ~(i=1,2)
\label{Conjugatemomenta1}
\eeqa
whose explicit expressions are given as\\[-20pt]
\beqa
\left.
\BA{lr}
\eta _1
=& - i\hbar \sum_{n=1}^N \!
\left(
{\displaystyle
x_n \frac{\partial}{\partial x_n }
-y_n \frac{\partial}{\partial y_n }
}
\right) ~,\\
\\[-10pt]
\eta _2
=& - i\hbar \sum_{n=1}^N \!
\left(
{\displaystyle 
x_n \frac{\partial}{\partial y_n } 
+ y_n \frac{\partial}{\partial x_n }
}
\right) ~.
\EA
\right \}
\label{Conjugatemomenta2}
\eeqa 
The commutation relations between these collective variables 
become as follows:\\[-18pt]
\beqa
\BA{rl}
\BA{lr}
\left[ \phi _i , \phi _j \right] = 0~, ~
\left[ \eta _i , \phi _j \right]
= 
- i\hbar \!~{\displaystyle \frac{r^2}{r_0 ^2}}~\!
\delta _{ij}~, & \\
\EA
(i, j=1,2) ,~
\left[ \eta _1 , \eta _2 \right] = - 2 i\hbar l ~,
\EA
\label{Commutationrelations}
\eeqa
where $r^2$ and $l$  are defined, respectively as
\beqa
\BA{rl}
r^2 \equiv {\displaystyle \frac{1}{N}} 
\sum_{n=1}^N (x _n ^2 + y _n ^2 ) ~,~
l
\!\equiv\!
\sum_{n=1}^N 
( x _n p_{y_n}- y _n p_{x_n}) ~.
\EA
\label{r**2andAngularmomentum}
\eeqa 
The mean square distance of the nucleon 
to the center of the nucleus is $r^2$
and its equilibrium value is approximately 
$r_0 ^2$ appearing in Eq. 
(\ref{Collectivecoordinates}), 
namely $\langle r^2 \rangle =r_0 ^2$,
while
$l$ is the total angular momentum operator of the system 
in the original representation.
From the second equation of Eq. (\ref{Commutationrelations}), 
an approximate commutation relation is derived as
\beqa
[\eta _i , \phi _j ] \cong - i\hbar {\delta }_{ij}~.~ (i, j=1,2)
\label{Approximate commutation relations}
\eeqa
Thus, we can get approximate conjugate momenta $\eta _1$ and $\eta _2$ 
to  the collective coordinates $\phi_1$ and $\phi_2$.
However, as is shown from Eq. (\ref{Commutationrelations}), 
the R.H.S. of the second equation does not take the value 
$- i\hbar {\delta }_{ij}$ and that of the third one does not vanish.
Then from these facts, it follows that the variables 
$\phi_i$ and $\eta _i$ are not canonically conjugate to each other.
In the commutation relations
(\ref{Commutationrelations}),
we are allowed to consider $r^2$ as an intrinsic variable,
in the sense that $\frac{1}{N}$ is small.
The commutator of $r^2$ with $\eta_i$ is of the order of $\frac{1}{N}$.
Moreover, in Eq.
(\ref{Commutationrelations}),
$r^2$ appear divided by $r_0^2$ which is of order $N$,
so $\frac{r^2}{r_0^2} \!=\! 1$ to a very good approximation.

\newpage


\def\thesection{\arabic{section}}
\setcounter{equation}{0}
\renewcommand{\theequation}{\arabic{section}.
\arabic{equation}}     

\section{Exact canonically conjugate momenta}
 
~~~We will now derive exact canonically conjugate momenta to $\phi _i$.
For this purpose, following
\cite{Tamura.56}
first we introduce operators $\pi _i$ defined by\\[-20pt]
\beqa
\pi_i
\equiv
\frac{r_0 ^2}{2} 
\left( r^{-2} \eta _i + \eta _i r^{-2} \right) .~ (i=1,2)
\label{definitionofpi}
\eeqa\\[-16pt] 
Then, the operators $\phi _i$ and $\pi _i$ satisfy 
the commutation relations\\[-20pt]
\beqa
[ \pi _i , \phi _j ]
= 
- i\hbar {\delta}_{ij} ,~ (i, j=1,2) ,~~
[ \pi _1 , \pi _2 ]
=
- 2 i\hbar \!
 \left( \frac{r^2}{r_0 ^2} \right) ^{\!\!-2} \!\!
( l - {\bm L}_{\mbox{coll}})~,
\label{commu.rela.pi.phi.pi.pi.}
\eeqa\\[-16pt] 
where ${\bm L}_{\mbox{coll}}$ is defined as\\[-26pt]
\beqa
{\bm L}_{\mbox{coll}} \equiv 2(\phi _1 \pi _2
- \phi _2 \pi _1 )~.
\label{collectiveangularmomentumoperator}
\eeqa\\[-16pt]
The commutation relations
(\ref{commu.rela.pi.phi.pi.pi.})
have no $\phi$ or $\phi\pi$ terms in each R.H.S.
and are very simple compared with those in I.
Thus the present exact canonical momentum $\pi _i$ is not self-recursive,
as opposed to the fact that 
the exact canonical momentum $\Pi _i$ in I is self-recursive,
using the discrete integral equation.
The term ${\bm L}_{\mbox{coll}}$
(\ref{collectiveangularmomentumoperator})
also appears in the Miyazima-Tamura's
collective description of a two-dimensional nucleus
\cite{MiyaTamu.56}.
If the inhomogeneous term in the R.H.S. of 
the second equation of
(\ref{commu.rela.pi.phi.pi.pi.})
disappears, 
the operators $\pi _i$ become independent from each other,
i.e., $[ \pi _1 , \pi _2 ] =0$
under the convecntion to conceive $r^{2}$
as constant $c$-number.
We can prove on the commutativity of  
the operator $\pi _i$ by requiring that
the inhomogeneous term vanishes.
Then, it turns out that $\pi _i$ are $exact$ canonical
conjugate to $\phi _i$.
Due to this fact and the form of  
(\ref{collectiveangularmomentumoperator}), 
the quantity ${\bm L}_{\mbox{coll}}$ can be regarded as 
a collective angular momentum operator.
Therefore, it is concluded that 
the operators $\{ \phi _i,~\pi _i \}$ is 
a set of $exact$ canonical variables, 
if we restrict the Hilbert space to 
the collective subspace $\ket{\mbox{coll.subspace}}$ 
which satisfies the subsidiary condition\\[-20pt] 
\beqa
(l - {\bm L}_{\mbox{coll}}) 
\ket{\mbox{coll.subspace}} = 0 ,~
\left[  \pi _1, \pi _2 \right] 
\ket{\mbox{coll.subspace}} = 0 .
\label{subsidiarycondition}
\eeqa\\[-20pt] 
This condition is very reasonable and also implies that 
our collective variables are the good ones 
in the collective subspace.
This kind of subsidiary condition also appears in I.

$\!\!\!\!$The old variable $\eta _i$ 
and the new one $\pi _i$
(\ref{definitionofpi}) 
are mutually related with $\pi _i$ and $\eta _i$, with $\phi _i$ as\\[-20pt]
\beqa
\eta _i = r_0 ^{-2} r^{2}\pi _i 
- i\hbar 2 r_0 ^2 r^{-2} \phi _i ,~~
\pi _i = r_0 ^2 r^{-2}\eta _i 
+ i\hbar 2 r_0 ^4 r^{-4} \phi _i ,~(i=1,2)
\label{newoldrelation}
\eeqa\\[-20pt]
where we have used the relations\\[-18pt]
\beqa
\left[ \eta _i , r^{2n} \right] 
= - i\hbar 4n r_0 ^2 r^{2(n-1)} \phi _i  ,~~
\left[ \pi _i , r^{2n} \right] 
= - i\hbar 4n r_0 ^4 r^{2(n-2)} \phi _i , ~ (i \!=\! 1,2) .
\label{etapirelations}
\eeqa\\[-16pt] 
The commutation relations between the variables $\eta_{i}$ and a kinetic operator $T$
are calculated as\\[-18pt]  
\beqa
\BA{lr}
\left[ \eta _1 , T \right] 
=
-i\hbar {\displaystyle \frac{\hbar ^2}{\mu}} \!
\sum_{n=1}^N \!
\left( \!
{\displaystyle  
\frac{\partial ^2}{\partial x_n ^2} 
- \frac{\partial ^2}{\partial y_n ^2}
} \!
\right) , ~
\left[  \eta _2  , T \right]
= - i\hbar {\displaystyle \frac{\hbar ^2}{\mu}} \!
\sum_{n=1}^N  \!
\left( \!
{\displaystyle  
\frac{\partial}{\partial  x_n} \frac{\partial}{\partial  y_n}
+ \frac{\partial}{\partial  y_n} \frac{\partial}{\partial  x_n}
} \!
\right) ,
\EA
\label{etaTcommurelations}
\eeqa\\[-10pt]
and the commutator
$\left[ T , r^{-2} \right]$
is given as\\[-16pt]
\beqa
\BA{ll}
\left[ T , r^{-2} \right]
= 
{\displaystyle \frac{2\hbar ^2}{\mu N r^4}} \!
\sum_{n=1}^N \! 
\left( \!
{\displaystyle  
x_n \frac{\partial}{\partial x_n }
+ y_n \frac{\partial}{\partial y_n } \!
}
\right) 
+ {\displaystyle \frac{2\hbar ^2}{\mu r^4 }} \!
\left( \! 1 - {\displaystyle \frac{2}{N}} \! \right) ,
\EA
\label{Tr2commurelation}
\eeqa\\[-14pt]
together with the commutators below\\[-16pt]  
\beqa
\left.
\BA{lr}
\left[ \left[ \eta_i , T \right] , r^{-2} \right] 
=
- {\displaystyle \frac{4\hbar ^2}{\mu N r^4}} \eta_i
- i\hbar {\displaystyle \frac{16\hbar ^2 r_0 ^2}{\mu N r^6}} \phi _i  , \\
\\[-12pt]
\left[ \left[ T , r^{-2} \right] ,  \eta_i  \right] 
=
-
i\hbar {\displaystyle \frac{16\hbar ^2 r_0 ^2}{\mu N  r^6}} \phi _i \!
\sum_{n=1}^N \!
\left( \!
{\displaystyle 
x_n \frac{\partial}{\partial x_n} 
+ y_n \frac{\partial}{\partial y_n} \!
}
\right) 
-
i\hbar {\displaystyle \frac{16\hbar ^2  r_0 ^2}{\mu  r^6}} \!
\left( \! 1 - {\displaystyle \frac{2}{N}} \! \right) \phi _i  .
\EA
\right\}
\label{Tr2commurelations}
\eeqa\\[-10pt]
We here briefly discuss the algebraic foundations of the present collective theory.
The set of nine operators
\{$T$ (2.1), $\phi_i$ (2.2), $\eta_i$ (2.4) and $[\eta_i, T]$ (3.7) (i\!=\!1, 2),
$r^2$ and $l$ (2.6)\}
without multiplication factors and the operator
$\sum_n (x_n p_{x_n} \!+\! y_n p_{y_n})$
span the 10 dimensional symplectic algebra $sp(2,R)$.
All of these ten generators and the commutator
(\ref{Tr2commurelation})
and the two commutators
(\ref{Tr2commurelations})
play important roles
in the next Section to derive a collective Hamiltonian.

\newpage


\def\thesection{\arabic{section}}
\setcounter{equation}{0}
\renewcommand{\theequation}{\arabic{section}.
\arabic{equation}}     

\section{$\pi _i$- and $\phi_i$- dependence of kinetic part of Hamiltonian}
 
\subsection{\large {\bf $\pi _i$-dependence of kinetic part of Hamiltonian}}

~~~~~In this subsection,
a remaining task is to express the kinetic part $T$ of the Hamiltonian in terms of
$\phi _i$ and $\pi_i$.
Along the same procedure as the one in I,
we first investigate $\pi _i$-dependence of $T$.
For this purpose,
we expand it in a power series of the exact canonical conjugate momenta $\pi _i$
as follows:\\[-20pt]
\beqa
\begin{array}{c}
T
=
T^{(0)}(\phi ; r^2) 
+ \sum^{2}_{i \!=\! 1} T_i^{(1)} (\phi ; r^2 ) \pi _i
+ \sum^{2}_{i,j \!=\! 1} T_{ij}^{(2)} (\phi ; r^2 ) \pi _i r^{-2} \pi _j r^2
+ \cdots~,
\end{array}
\label{Tphiexpansion}
\eeqa\\[-18pt]
where $T_{ij}^{(n)} (\phi ; r^2 )$
are unknown expansion coefficients.
In order to get the explicit expression for $T^{n}(n \neq 0)$,
using the commutation relation
$[ \pi _i , \phi _j ] \!=\! - i\hbar {\delta}_{ij}$,
we take the commutators with $\phi _i$
in the following way:\\[-20pt]
\beqa
\left.
\BA{lr}
\left[ T , \phi _i \right] 
=
- i\hbar T_i^{(1)} (\phi ; r^2 ) ,~(i=1,2) \\
\\[-8pt]
\left[ \left[ T , \phi _i \right] , \phi _j \right] 
= (-i\hbar) ^2 2 T_{ij}^{(2)} (\phi ; r^2 ) ,~(i, j=1,2) \\
\\[-8pt]
\left[ \left[ \left[ T , \phi _i \right] 
, \phi _j \right] , \phi _k \right]
= (-i\hbar) ^3 6 T_{ijk}^{(2)} (\phi ; r^2 ) ,~(i, j, k=1,2) , 
\\[-2pt]
\vdots .
\EA
\right\}
\label{Tcommutatorswithphi1}
\eeqa\\[-14pt]
We can easily calculate the L.H.S. of
(\ref{Tcommutatorswithphi1})
by making explicit use of the definitions
(\ref{Collectivecoordinates})
and
(\ref{definitionofpi})
and by taking commutators with $\phi _i$ successively:\\[-20pt]
\beqa
\left.
\BA{lr}
\left[ T , \phi _i \right] 
=
- i\hbar {\displaystyle \frac{1}{\mu N r_0 ^2 }} \eta _i
=
- i\hbar {\displaystyle \frac{1}{2 \mu N r_0 ^4 }} 
(r^2 \pi _i + \pi _i r^2 ) ,~(i=1,2) \\
\\[-14pt]
\left[ \left[ T , \phi _i \right] , \phi _j \right] 
= (i\hbar) ^2 {\displaystyle \frac{r^2}{\mu N r_0 ^4 }} 
\delta _{ij} ,~(i, j \!=\!  \!1\! ,2) \\
\\[-8pt]
\left[ \left[ \left[ T , \phi _i \right] 
, \phi _j \right] , \phi _k \right] = 0 ,~(i, j, k \!=\! 1,2) 
\\[-2pt]
\vdots .
\EA
\right\}
\label{Tcommutatorswithphi2}
\eeqa\\[-14pt]
Comparing
(\ref{Tcommutatorswithphi2})
with
(\ref{Tcommutatorswithphi1}), 
the $T^{(n)}$ are determined as\\[-18pt]
\beqa
\left.
\BA{lr}
T_i ^{(1)} (\phi ; r^2 ) = 0 , \\
\\[-12pt]
T_{ij}^{(2)} (\phi ; r^2 )
= 
T_{ji}^{(2)} (\phi ; r^2 ) 
=
{\displaystyle \frac{r^2 }{2\mu N r_0 ^4 }} \delta _{ij} , \\
\\[-12pt]
T^{(n)}(\phi ; r^2 ) = 0 ,~(n \ge 3)
\EA
\right\}
\label{Tndetermined}
\eeqa\\[-10pt]
in which it should be noted that
$T^{(2)}$ has no $\phi$ term.
This result brings the essential difference from the corresponding one in I. 
Substituting
(\ref{Tndetermined})
into
(\ref{Tphiexpansion})
and using the second relation of
(\ref{etapirelations}),
we can get the exact $\pi_i$-dependence of the kinetic part $T$
of Hamiltoniasn as follows:\\[-20pt]
\beqa
\BA{lr}
T
=
T^{(0)}(\phi ; r^2 ) 
+
{\displaystyle \frac{1}{2\mu N r_0 ^4}} \!
\sum^{2}_{i \!=\! 1} r^2 \pi _j r^{-2} \pi _i r^2 \\
\\[-12pt]
~~~
=
T^{(0)} (\phi ; r^2 ) 
-
{\displaystyle \frac{4\hbar ^2 }{\mu  N r^2}}
+
{\displaystyle \frac{r^2 }{2\mu N r_0 ^4}} \!
\sum^{2}_{i \!=\! 1} \pi _i \pi _i 
+
{\displaystyle \frac{16\hbar ^2 r_0 ^4}{\mu N r^6}} \!
\sum^{2}_{i \!=\! 1} \phi _i \phi _i 
-
i\hbar {\displaystyle \frac{2}{\mu N r^2}} \!
\sum^{2}_{i \!=\! 1} \phi _i \pi _i  ,
\EA
\label{Tndeterminedpi}
\eeqa\\[-14pt]
in which the third term is the kinetic energy part of
the two-dimensional collective motion in nuclei
given by Miyazima-Tamura
\cite{MiyaTamu.56}. 
The last two terms do not appear in Miyazima-Tamura's work
because they adopt the canonical transformation method together
with the subsidiary condition.
These terms are regarded as a harmonic potential and
an anharmonic self-coupling, respectively,
for each collective motion.

Finally,
we should stress the fact that up to this stage,
all the expressions are exact.

\subsection{\large {\bf $\phi _i$-dependence of kinetic part of Hamiltonian}}

~~~Our next task is to determine the term
$T^{(0)} (\phi ;  r^2 )$.
For this purpose,
we expand it in a power series of the collective coordinates
$\phi _j$ in the form\\[-20pt]
\beqa
\!\!\!\!
\begin{array}{c}
T^{(0)}(\phi ; r^2) 
\!=\!
C_{0}(r^{2})
\!+\! \sum^{2}_{i \!=\! 1} \! C_{1i}(r^{2}) \phi _i
\!+\! \sum^{2}_{i,j \!=\! 1} \! C_{2ij}(r^{2}) \phi _i \phi _j
\!+\! \sum^{2}_{i,j,k \!=\! 1} \! C_{3ijk}(r^{2}) \phi _i \phi _j \phi _k
\!+\! \cdots ,
\end{array}
\label{T0phiexpansion}
\eeqa\\[-16pt]
where
$C_{2ij}(r^{2}) \!=\! C_{2ji}(r^{2})$, etc.
In the above,
the expansion coefficients
$C_{n}(r^{2})$
are determined in a manner quite similar to the one used before.

First, using
(\ref{Tndeterminedpi}),
we have the commutation relation between
the operator $\pi _i$ and $T^{(0)} (\phi ;  r^2 )$
as\\[-24pt]
\beqa
\BA{lr}
\left[ \pi _i , T^{(0)} (\phi ;  r^2 ) \right]
=
\left[ \pi _i , T \right] 
+
i\hbar {\displaystyle \frac{48\hbar ^2 r_0 ^4}{\mu N r^6 }} \phi _i 
- 
i\hbar {\displaystyle \frac{192\hbar ^2 r_0 ^8}{\mu N r^{10}}} \phi _i \!
\sum^{2}_{j \!=\! 1} \phi _j \phi _j \\
\\[-12pt]
~~~~~~~~~~~~~~~~~~~~~~
+
{\displaystyle \frac{2\hbar ^2}{\mu N r^2}} \pi _i
-
{\displaystyle \frac{8\hbar ^2 r_0 ^4}{\mu N r^6}} \phi _i \!
\sum^{2}_{j \!=\! 1} \phi _j \pi _j
+
i\hbar {\displaystyle \frac{2}{\mu N r^2 }} \phi _i  \!
\sum^{2}_{j \!=\! 1} \pi _j \pi _j  ,~(i=1,2)
\EA
\label{commupiT0}
\eeqa\\[-12pt]
in which we have used  the first relation of
(\ref{commu.rela.pi.phi.pi.pi.})
and the second relation of
(\ref{etapirelations}).
The first commutator
$\left[ \pi _i , T \right]$
in
(\ref{commupiT0})
is computed as\\[-20pt]
\beqa
\!\!\!\!\!\!
\left.
\BA{lr}
\left[ \pi _i , T \right] 
\!=\!
{\displaystyle \frac{r_0 ^2}{2}} \!
\left[ r^{-2} \eta _i  \!+\!  \eta _i r^{-2}, T \right]
\!=\!
r_0 ^2 r^{-2} \! \left[ \eta _i , T \right]
\!-\!
r_0 ^2 \! \left[ T, r^{-2} \right] \! \eta _i
\!+\!
{\displaystyle \frac{r_0 ^2}{2}} \!
\left[ \left[ \eta _i , T \right],  r^{-2} \right]
\!+\!
{\displaystyle \frac{r_0 ^2}{2}} \!
\left[ \left[ T , r^{-2} \right],  \eta _i \right] \\
\\[-10pt]
=
r_0 ^2 r^{-2} \!
\left[  \eta_i , T \right] 
\!-\!
{\displaystyle \frac{\hbar ^2 r_0 ^2}{2\mu}} \!\!
\left\{ \!\!
{\displaystyle \frac{4}{N}} r^{-4} \!
\sum_{n=1}^N \!\!
\left( \!\!
{\displaystyle 
x_n \frac{\partial}{\partial x_n} 
\!+\!
y_n \frac{\partial}{\partial y_n}
} \!\!
\right) \!
\eta _i 
\!+\!
i\hbar {\displaystyle \frac{16}{N}} r_0 ^2 r^{-6} \phi _i \!
\sum_{n=1}^N \!\!
\left( \!\!
{\displaystyle 
x_n \frac{\partial}{\partial x_n} 
\!+\!
y_n \frac{\partial}{\partial y_n}
} \!\!
\right) 
\right. \\
\\[-10pt]
\left.
~~
+ 4
\left( \! 1 - {\displaystyle \frac{1}{N}} \! \right) \!
r^{-4} \eta _i 
+ i\hbar 16
\left( \! 1 - {\displaystyle \frac{1}{N}} \! \right) \!
r_0 ^2 r^{-6} \phi _i  \!
\right \} ,~ (i=1,2)
\EA
\right.
\label{commutatorpaiT}
\eeqa\\[-12pt] 
where we have used the commutators
(\ref{Tr2commurelation})
and
(\ref{Tr2commurelations}).
Substituting
(\ref{commutatorpaiT})
into
(\ref{commupiT0})
and the second equation of
(\ref{newoldrelation}),
we can get
$\left[ \pi _i  , T^{(0)} (\phi ; r^2) \right]$
as\\[-16pt]
\beqa
\left.
\BA{lr}
\left[ \pi _i  , T^{(0)} (\phi ; r^2) \right]  \\
\\[-10pt]
=
r_0 ^2 r^{-2} \!
\left[  \eta_i , T \right] 
\!-\!
{\displaystyle \frac{\hbar ^2 r_0 ^2}{2\mu}} \!\!
\left\{ \!\!
{\displaystyle \frac{4}{N}} r^{-4} \!
\sum_{n=1}^N \!\!
\left( \!\!
{\displaystyle 
x_n \frac{\partial}{\partial x_n} 
\!+\!
y_n \frac{\partial}{\partial y_n}
} \!\!
\right) \!
\eta _i 
\!+\!
i\hbar {\displaystyle \frac{16}{N}} r_0 ^2 r^{-6} \phi _i \!
\sum_{n=1}^N \!\!
\left( \!\!
{\displaystyle 
x_n \frac{\partial}{\partial x_n} 
\!+\!
y_n \frac{\partial}{\partial y_n}
} \!\!
\right) 
\right. \\
\\[-10pt]
\left.
~~
+ 4
\left( \! 1 - {\displaystyle \frac{1}{N}} \! \right) \!
r^{-4} \eta _i 
+ i\hbar 16
\left( \! 1 - {\displaystyle \frac{1}{N}} \! \right) \!
r_0 ^2 r^{-6} \phi _i  \!
\right \} \\
\\[-10pt]
+
{\displaystyle \frac{{\hbar}^{2} r_0 ^2}{2\mu}} \!
\left \{ \!
i\hbar {\displaystyle \frac{96}{N}} r_0 ^2 r^{-6} \phi _i  
\!-\!
i\hbar {\displaystyle \frac{384}{N}} r_0 ^6 r^{-10} \phi _i 
\sum^{2}_{j \!=\! 1} \phi _j  \phi _j
\!+\!
{\displaystyle \frac{4}{N} } r^{-4} \eta _i
\!+\!
i\hbar {\displaystyle \frac{8}{N}} r_0 ^2 r^{-6} \phi _i
\right. \\
\\[-10pt]
\left.
-
{\displaystyle \frac{16}{N}} r_0 ^2 r^{-6}
\phi _i \!
\sum^{2}_{j \!=\! 1} \phi _j  \pi _j \!
\right \}
\!+\!
i \hbar {\displaystyle \frac{2}{\mu N r^2}} \phi _i \!
\sum^{2}_{j \!=\! 1} \pi _j  \pi _j  .~ (i=1,2)
\EA
\right.
\label{commutatorpaiT0}
\eeqa\\[-12pt] 
To carry further computation of
(\ref{commutatorpaiT0}),
the following approximate relations play crucial roles:\\[-18pt]
\beqa
\!\!\!\!\!\!\!\!\!
\BA{lr}
\sum_{n=1}^N \!\!
\left( \!
{\displaystyle 
x_n \frac{\partial}{\partial x_n} 
\!+\!
y_n \frac{\partial}{\partial y_n}
} \!
\right) 
\eta _i 
\!\approx\!
\eta _i
\!+\!
f(N) \eta _i 
\!+\!
{\displaystyle \frac{\mu}{ \hbar^2} \frac{1}{2}} N \! r^{2}
\left[  \eta_i , T \right] 
\!-\!
i \hbar N \! r^{2}_{0} \phi _i \!
\sum_{n=1}^N \!
\left( \!
{\displaystyle 
\frac{\partial^{2}}{\partial x_n^{2}} 
\!+\!
\frac{\partial^{2}}{\partial y_n^{2}}
} \!
\right) \! ,
\EA
\label{relationeta}
\eeqa\\[-32pt]
\beqa
\BA{c}
\sum^{2}_{j \!=\! 1} \phi _j  \pi _j
\!\approx\!
- i \hbar
{\displaystyle \frac{1}{2}} \!
\sum_{n=1}^N \!\!
\left( \!
{\displaystyle 
x_n \frac{\partial}{\partial x_n} 
+
y_n \frac{\partial}{\partial y_n}
} \!
\right) \!
\!+\!
i \hbar 2 r^{4}_{0} r^{-4} \!
\sum^{2}_{j \!=\! 1} \phi _j  \phi _j ,
\EA
\label{sumphipi}
\eeqa\\[-30pt]
\beqa
\BA{c}
\sum^{2}_{j \!=\! 1} \pi _j  \pi _j
\!\approx\!
\hbar^{2} 4 r^{4}_{0} r^{-4}
\!-\! 
\hbar^{2} 20 r^{8}_{0} r^{-8} \!
\sum^{2}_{j \!=\! 1} \phi _j  \phi _j \\
\\[-8pt]
+
\hbar^{2} 2 r^{4}_{0} r^{-4} \!
\sum_{n=1}^N \!\!
\left( \!
{\displaystyle 
x_n \frac{\partial}{\partial x_n} 
+
y_n \frac{\partial}{\partial y_n}
} \!
\right) \!
\!-\!
\hbar^{2} N r^{4}_{0} r^{-2} \!
\sum_{n=1}^N \!\!
\left( \!
{\displaystyle 
\frac{\partial^{2}}{\partial x_n^{2}} 
\!+\!
\frac{\partial^{2}}{\partial y_n^{2}}
} \!
\right) \! ,
\EA
\label{sumpipi}
\eeqa\\[-12pt]
where we we have used another approximate relation and
an approximate mean-value\\[-14pt]
\beqa
\BA{c}
\sum^{2}_{j \!=\! 1} \phi _j  \eta _j
\!\approx\!
- i \hbar
{\displaystyle \frac{1}{2}}
r^{-2}_{0} r^{2} \!
\sum_{n=1}^N \!\!
\left( \!
{\displaystyle 
x_n \frac{\partial}{\partial x_n} 
\!+\!
y_n \frac{\partial}{\partial y_n}
} \!
\right) \! ,~
\sum_{n=1}^N
{\displaystyle 
\langle
x_n \frac{\partial}{\partial x_n} 
\!+\!
y_n \frac{\partial}{\partial y_n}
\rangle
}
\!=\!
f(N) ,
\EA
\label{sumphietameanvalue}
\eeqa\\[-14pt]
where the average $\langle \cdot \rangle$ is taken
on the collective subspace
$\ket{\mbox{coll.subspace}}$.
The unknown function $f(N)$ is determined later.
A derivation of equation
(\ref{relationeta})
is given in detail in Appendix A.
Substititing
(\ref{relationeta}), (\ref{sumphipi})
and
(\ref{sumpipi})
into
(\ref{commutatorpaiT0}),
we reach the following final result:\\[-16pt]
\beqa
\BA{lr}
\left[ \pi _i  , T^{(0)} (\phi ; r^2) \right]  \\
\\[-12pt]
\!=\!
r_0 ^2 r^{-2} \!
\left[  \eta_i , T \right] 
\!-\!
{\displaystyle \frac{\hbar ^2 r_0 ^2}{2\mu}} \!\!
\left[ \!\!
{\displaystyle \frac{4}{N}} r^{-4} \!
\left\{ \!
\eta _i
\!+\!
f(N) \eta _i 
\!+\!
{\displaystyle \frac{\mu}{ \hbar^2} \frac{1}{2}} N \! r^{2}
\left[  \eta_i , T \right] 
\!-\!
i \hbar N \! r^{2}_{0} \phi _i \!
\sum_{n=1}^N \!
\left( \!
{\displaystyle 
\frac{\partial^{2}}{\partial x_n^{2}} 
\!+\!
\frac{\partial^{2}}{\partial y_n^{2}}
} \!
\right) \!\!
\right\}
\right. \\
\\[-12pt]
\left.
+
i\hbar {\displaystyle \frac{16}{N}} r_0 ^2 r^{-6} \phi _i \!
\sum_{n=1}^N \!\!
\left( \!
{\displaystyle 
x_n \frac{\partial}{\partial x_n} 
\!+\!
y_n \frac{\partial}{\partial y_n}
} \!
\right)
\!+\!
4 \!
\left( \! 1 - {\displaystyle \frac{1}{N}} \! \right) \!
r^{-4} \eta _i 
\!+\!
i\hbar
16 \!
\left( \! 1 - {\displaystyle \frac{1}{N}} \! \right) \!
r_0 ^2 r^{-6} \phi _i  \!
\right] \\
\\[-12pt]
+
{\displaystyle \frac{{\hbar}^{2} r_0 ^2}{2\mu}} \!
\left[ \!
i\hbar {\displaystyle \frac{96}{N}} r_0 ^2 r^{-6} \phi _i  
\!-\!
i\hbar {\displaystyle \frac{384}{N}} r_0 ^6 r^{-10} \phi _i 
\sum^{2}_{j \!=\! 1} \phi _j  \phi _j
\!+\!
{\displaystyle \frac{4}{N} } r^{-4} \eta _i
\!+\!
i\hbar {\displaystyle \frac{8}{N}} r_0 ^2 r^{-6} \phi _i
\right. \\
\\[-12pt]
\left.
-
{\displaystyle \frac{16}{N}} r_0 ^2 r^{-6}
\phi _i \!
\left\{
- i \hbar
{\displaystyle \frac{1}{2}} \!
\sum_{n=1}^N \!\!
\left( \!
{\displaystyle 
x_n \frac{\partial}{\partial x_n} 
+
y_n \frac{\partial}{\partial y_n}
} \!
\right) \!
\!+\!
i \hbar 2 r^{4}_{0} r^{-4} \!
\sum^{2}_{j \!=\! 1} \phi _j  \phi _j \!
\right\} \!
\right] \\
\\[-12pt]
\!+\!
i \hbar {\displaystyle \frac{2}{\mu N r^2}} \phi _i \!
\left\{ \!
\hbar^{2} 4 r^{4}_{0} r^{-4}
\!-\! 
\hbar^{2} 20 r^{8}_{0} r^{-8} \!
\sum^{2}_{j \!=\! 1} \phi _j  \phi _j
\!-\!
\hbar^{2} N r^{4}_{0} r^{-2} \!
\sum_{n=1}^N \!\!
\left( \!
{\displaystyle 
\frac{\partial^{2}}{\partial x_n^{2}} 
\!+\!
\frac{\partial^{2}}{\partial y_n^{2}}
} \!
\right)
\right. \\
\\[-12pt]
\left.
+
\hbar^{2} 2 r^{4}_{0} r^{-4} \!
\sum_{n=1}^N \!\!
\left( \!
{\displaystyle 
x_n \frac{\partial}{\partial x_n} 
+
y_n \frac{\partial}{\partial y_n}
} \!
\right) \!\!
\right\} \\
\\[-12pt]
= 
-
{\displaystyle \frac{\hbar ^2 r_0 ^2}{2\mu}}
4 r^{-4} \!
\left\{ \!
{\displaystyle \frac{1}{N}} f(N)
\!+\!
\left( \! 1 - {\displaystyle \frac{1}{N}} \! \right) \!
\right\}
\eta _i  \\
\\[-12pt]
-
{\displaystyle \frac{\hbar ^2 r_0 ^2}{2\mu}}
i\hbar \!
\left\{ \!
16 \!
\left( \! 1 \!-\! {\displaystyle \frac{1}{N}} \! \right) \!
r_0 ^2 r^{-6} \phi _i 
\!-\!
{\displaystyle \frac{120}{N}} r_0 ^2 r^{-6} \phi _i  
\!+\!
{\displaystyle \frac{496}{N}} r_0 ^6 r^{-10} \phi _i \!
\sum^{2}_{j \!=\! 1} \phi _j  \phi _j \!
\right\}  .~ (i=1,2)
\EA
\label{commutatorpaiT02}
\eeqa\\[-12pt]
The double commutator
$
\left[ \left[ \pi_i, T^{(0)} \! (\phi ;r^2) \right], 
\phi_j \right]
\!=\!
0
$
is indispensable for our theory.
To achieve the commutability,
it is necessary that
$\left[ \pi_i, T^{(0)} \! (\phi ;r^2) \right]$
depends only on the variable $\phi$.
For this aim, we demand that
the first term $\eta _i$ of the second line from the bottom in
(\ref{commutatorpaiT02})
vanish on the
$\ket{\mbox{coll.subspace}}$.
Then, $f(N)$ is settled as
$
f(N)
\!\!=\!\!
-
N \!+\! 1
$. 
As mentioned before,
the present approach
essentially lies on the symplectic algebra $sp(2,R)$.
In the ten generators of the algebra,
in particular, here
the generator
$\sum_n \! (x_n p_{x_n} \!\!+\! y_n p_{y_n})$
is contracted as
$\sum_n \! \langle x_n p_{x_n} \!\!+\! y_n p_{y_n} \rangle$
(\ref{sumphietameanvalue}). 
The present approach contrasts with the procedure followed
in the contracted symplectic model
(Castan\"{o}s and Draayer
\cite{CastanosDraayer.89})
in which all the generators are contracted.
Using
(\ref{commutatorpaiT02}), 
$
f(N)
\!=\!
-
N \!+\! 1
$,
the first of
(\ref{commu.rela.pi.phi.pi.pi.})
and the second of
(\ref{subsidiarycondition}),
and for the sake of simplicity
discarding the contributions from effects by the terms
$\left[ \pi_i , r^{-6} \right]$
and
$\left[ \pi_i , r^{-10} \right]$,
we get the following commutation relations:\\[-16pt]
\beqa
\!\!\!\!\!\!
\BA{lr}
\left[ \pi _i , \left[ \pi _j , T^{(0)} (\phi ; r^2) \right] \right]
\!=\! 
-
{\displaystyle \frac{\hbar ^2 r_0 ^2}{2\mu}}
\hbar ^2 \!
\left\{ \!
16 \!
\left( \! 1 \!-\! {\displaystyle \frac{1}{N}} \! \right) \!
r_0 ^2 r^{-6} \delta_{i j} 
\!+\! 
{\displaystyle \frac{4}{N}}
r_0 ^2 r^{-6} \delta_{i j} 
\!+\!
2 {\displaystyle \frac{496}{N}} r_0 ^6 r^{-10} 
\phi _i \phi _j \!
\right\} \! ,
\EA
\label{commutatorpaipaiT0}
\eeqa\\[-24pt]
\beqa
\!\!\!\!\!\!
\BA{lr}
\left[ \pi _i , \left[ \pi _j , \left[ \pi _k , T^{(0)} (\phi ; r^2) \right] \right] \right]
\!=\! 
{\displaystyle \frac{\hbar ^2 r_0 ^2}{2\mu}}
i\hbar
\hbar ^2 
2 {\displaystyle \frac{496}{N}} r_0 ^6 r^{-10} \!
\left( 
\delta_{i j} \phi _k
\!+\!
\phi _j \delta_{i k} 
\right)  ,
\EA
\label{commutatorpaipaipaiT0}
\eeqa\\[-24pt]
\beqa
\!\!\!\!\!\!
\BA{lr}
\left[ \pi _i , \left[ \pi _j , \left[ \pi _k , \left[ \pi _l , T^{(0)} (\phi ; r^2)
\right] \right] \right] \right]
\!=\! 
{\displaystyle \frac{\hbar ^2 r_0 ^2}{2\mu}}
\hbar ^4 
2 {\displaystyle \frac{496}{N}} r_0 ^6 r^{-10} \!
\left( 
\delta_{jk} \delta_{i l} 
\!+\!
\delta_{i k} \delta_{jl} 
\right)  ,
\EA
\label{commutatorpaipaipaipaiT0}
\eeqa\\[-24pt]
\beqa
\BA{lr}
\left[ \pi _i \left[ \pi _j , \left[ \pi _k , \left[ \pi _l , \left[ \pi _m , T^{(0)} (\phi ; r^2)
\right] \right] \right] \right] \right]
\!=\! 0 ,
(i,j,k,l,m \!=\! 1,2) .
\EA
\label{commutatorpaipaipaipaipaiT0}
\eeqa\\[-12pt]
In the derivation of
(\ref{commutatorpaipaiT0}),
we have used the approximate relation
$
\sum^{2}_{i \!=\! 1} \phi _i  \phi _i
\!\approx\!
{\displaystyle \frac{1}{4}}
r_0 ^{-4} r^{4}
$.

\newpage

By a similar procedure to the previous one,
we take the commutation relations
between
$T^{(0)} (\phi ; r^2)$
expanded as
(\ref{T0phiexpansion})
with $\pi _i$:\\[-14pt]
\beqa
\!\!\!\!\!\!\!\!\!\!\!\!
\left.
\BA{lr}
\left[ \pi _i, T^{(0)} (\phi ; r^2) \right] 
\!=\! 
- i\hbar C_{i}(r^2)
\!-\!
2 i\hbar \!
\sum^{2}_{j \!=\! 1} \!
C_{2ij}(r^2)  \phi _j
\!-\!
3 i\hbar \!
\sum^{2}_{j,k \!=\! 1} \!
C_{3ijk}(r^2)  \phi _j \phi _k
\!+\!
\cdots , \\
\\[-4pt]
\left[ \pi _i , \left[ \pi _j , T^{(0)} (\phi ; r^2) \right] \right]
\!=\! 
-
2 \hbar^2 \!\!
\left\{ \!
C_{2ij}(r^2)
\!+\!
3 \!
\sum^{2}_{k \!=\! 1} \!
C_{3ijk}(r^2)  \phi _k
\!+\!
6 \!
\sum^{2}_{k,l \!=\! 1} \!
C_{4ijkl}(r^2)  \phi _k \phi _l \!
\right\} \!
\!+\!
\cdots \! , \\
\\[-4pt]
\left[ \pi _i , \left[ \pi _j , \left[ \pi _k , T^{(0)} (\phi ; r^2) \right] \right] \right]
\!=\! 
6 i\hbar \hbar^2 \!
\left\{
C_{3ijk}(r^2)
\!+\!
4 \!
\sum^{2}_{l \!=\! 1} \!
C_{4ijkl}(r^2)  \phi _l
\right\} 
\!+\!
\cdots , \\
\\[-4pt]
\left[ \pi _i , \left[ \pi _j , \left[ \pi _k , \left[ \pi _l , T^{(0)} (\phi ; r^2)
\right] \right] \right] \right]
\!=\! 
24 \hbar^4
C_{4ijkl}(r^2) 
\!+\!
\cdots , \\
\\[-4pt]
\left[ \pi _i , \left[ \pi _j , \left[ \pi _k , \left[ \pi _l , \left[ \pi _m , T^{(0)} (\phi ; r^2)
\right] \right] \right] \right] \right]
\!=\! 
0 ,
(i,j,k,l,m \!=\! 1,2) .
\EA \!\!\!
\right\}
\label{commutatorpaiT0expa}
\eeqa
Comparing
(\ref{commutatorpaiT0expa})
with the last line of
(\ref{commutatorpaiT02})
and with equations from
(\ref{commutatorpaipaiT0})
to
(\ref{commutatorpaipaipaipaipaiT0})
and using the approximate relation
$
\sum^{2}_{i \!=\! 1} \phi _i  \phi _i
\!\approx\!
{\displaystyle \frac{1}{4}}
r_0 ^{-4} r^{4}
$, 
$C_{n}$ are determined as\\[-14pt]
\beqa
\left.
\BA{lr}
C_1 (r^2 ) = 0 , \\
\\[-8pt]
C_{2ij} (r^2 )
= 
{\displaystyle \frac{\hbar ^2 r_0 ^2}{2\mu}} \!
\left\{ \!
8 \!
\left( \! 1 \!-\! {\displaystyle \frac{1}{N}} \! \right) \!
r_0 ^2 r^{-6} \delta_{i j} 
\!+\! 
{\displaystyle \frac{2}{N}}
r_0 ^2 r^{-6} \delta_{i j} 
\!+\!
{\displaystyle \frac{496}{N}} r_0 ^6 r^{-10} 
\phi _i \phi _j \!
\right\} , \\
\\[-8pt]
C_{3ijk} (r^2 )
\!=\! 
{\displaystyle \frac{\hbar ^2 r_0 ^2}{2\mu}} 
{\displaystyle \frac{1}{3}}
{\displaystyle \frac{496}{N}} r_0 ^6 r^{-10} \!
\left( 
\delta_{i j} \phi _k
\!+\!
\phi _j \delta_{i k} 
\right) , \\
\\[-8pt]
C_{4ijkl} (r^2 )
\!=\! 
{\displaystyle \frac{\hbar ^2 r_0 ^2}{2\mu}} 
{\displaystyle \frac{1}{3}}
{\displaystyle \frac{124}{N}} r_0 ^6 r^{-10} \!
\left( 
\delta_{jk} \delta_{i l} 
\!+\!
\delta_{i k} \delta_{jl} 
\right) , \\
\\[-8pt]
C_{n} (r^2 )
\!=\! 
0 , ~(n \ge 5) .
\EA
\right\}
\label{Cndetermined}
\eeqa
Substituting
(\ref{Cndetermined})
into
(\ref{T0phiexpansion}),
we have\\[-14pt]
\beqa
\begin{array}{lr}
T^{(0)}(\phi ; r^2) 
=
C_{0}(r^{2}) \\
\\[-8pt]
\!+\!
\sum^{2}_{i,j \!=\! 1} \!
{\displaystyle \frac{\hbar ^2 r_0 ^2}{2\mu}} \!
\left\{ \!
8 \!
\left( \! 1 \!-\! {\displaystyle \frac{1}{N}} \! \right) \!
r_0 ^2 r^{-6} \delta_{i j} 
\!+\! 
{\displaystyle \frac{2}{N}}
r_0 ^2 r^{-6} \delta_{i j} 
\!+\!
{\displaystyle \frac{496}{N}} r_0 ^6 r^{-10} 
\phi _i \phi _j \!
\right\} \!
\phi _i \phi _j \\
\\[-8pt]
\!+\!
\sum^{2}_{i,j,k \!=\! 1} \!
{\displaystyle \frac{\hbar ^2 r_0 ^2}{2\mu}} 
{\displaystyle \frac{1}{3}}
{\displaystyle \frac{496}{N}} r_0 ^6 r^{-10} \!
\left( 
\delta_{i j} \phi _k
\!+\!
\phi _j \delta_{i k} 
\right) 
\phi _i \phi _j \phi _k \\
\\[-8pt]
\!+\!
\sum^{2}_{i,j,k,l \!=\! 1} \!
{\displaystyle \frac{\hbar ^2 r_0 ^2}{2\mu}} 
{\displaystyle \frac{1}{3}}
{\displaystyle \frac{124}{N}} r_0 ^6 r^{-10} \!
\left( 
\delta_{jk} \delta_{i l} 
\!+\!
\delta_{i k} \delta_{jl} 
\right)
\phi _i \phi _j \phi _k \phi _l
\!+\! \cdots , \\
\\[-8pt]
=
C_{0}(r^{2})\\
\\[-8pt]
\!+
{\displaystyle \frac{\hbar ^2 r_0 ^2}{2\mu}} \!
\left\{ \!
8 \!
\left( \! 1 \!-\! {\displaystyle \frac{1}{N}} \! \right) \!
r_0 ^2 r^{-6} \!
\sum^{2}_{i \!=\! 1} \! \phi _i  \phi _i 
\!+\! 
{\displaystyle \frac{2}{N}}
r_0 ^2 r^{-6} \!
\sum^{2}_{i \!=\! 1} \! \phi _i  \phi _i 
\!+\!
{\displaystyle \frac{496}{N}} r_0 ^6 r^{-10} \!
\sum^{2}_{i \!=\! 1} \! \phi _i  \phi _i \!
\sum^{2}_{j \!=\! 1} \! \phi _j  \phi _j \!
\right\} \\
\\[-8pt]
\!+\!
{\displaystyle \frac{\hbar ^2 r_0 ^2}{2\mu}} 
{\displaystyle \frac{2}{3}}
{\displaystyle \frac{496}{N}} r_0 ^6 r^{-10} \! 
\sum^{2}_{i \!=\! 1} \phi _i  \phi _i \!
\sum^{2}_{j \!=\! 1} \phi _j  \phi _j 
+
{\displaystyle \frac{\hbar ^2 r_0 ^2}{2\mu}} 
{\displaystyle \frac{2}{3}}
{\displaystyle \frac{124}{N}} r_0 ^6 r^{-10} \!
\sum^{2}_{i \!=\! 1} \phi _i  \phi _i \!
\sum^{2}_{j \!=\! 1} \phi _j  \phi _j
+
\cdots .
\end{array}
\label{T0phiexpansion2}
\eeqa\\[-8pt]
From
(\ref{T0phiexpansion2}), 
we obtain the final result
in the following form:
\beqa
\!\!\!\!\!\!
\begin{array}{cc}
T^{(0)}(\phi ; r^2) 
\!=\!
C_{0}(r^{2}) 
+
{\displaystyle \frac{\hbar ^2 r_0 ^{2}}{2\mu}} 
2 \!
\left( \! 4 \!-\! {\displaystyle \frac{3}{N}} \! \right) \!
r_0 ^2 r^{-6} \! 
\sum^{2}_{i \!=\! 1} \phi _i  \phi _i \!
+
{\displaystyle \frac{\hbar ^2 r_0 ^{2}}{2\mu}} ~\!
{\displaystyle \frac{2728}{3N}} ~\!
r_0 ^6 r^{-10} \!
\left( \sum^{2}_{i \!=\! 1} \phi _i  \phi _i \right)^{2} \! .
\end{array}
\label{T0phiexpansion3}
\eeqa\\[-16pt]
The new kinetic energy term
(\ref{T0phiexpansion3})
gives a strong anharmonicity in the vibrational regime.

\newpage

\subsection{\large {\bf Determination of constant term $C_{0}(r^{2})$} and
final expression for kinetic part  $T$ of Hamiltonian}
 
~~~In this Subsection,
 we determine the constant term
$C_{0}(r^{2}) $.
Using
(\ref{T0phiexpansion3})
and (\ref{Tndeterminedpi}),
$C_{0}(r^{2}) $
is given as\\[-20pt]
\beqa
\begin{array}{cc}
C_{0}(r^{2})  
\!=\!
T^{(0)}(\phi ; r^2)
\!-\!
{\displaystyle \frac{\hbar ^2 r_0 ^{2}}{2\mu}} \!
\left\{ \!
2 \!
\left( \! 1 \!-\! {\displaystyle \frac{1}{N}} \! \right)  
\!+\!
{\displaystyle \frac{1}{2}} 
{\displaystyle \frac{1}{N}}
\!+\!
{\displaystyle \frac{31}{N}} 
\!+\!
{\displaystyle \frac{5}{6}}
{\displaystyle \frac{31}{N}} \!
\right\} \!
r_0 ^{-2} r^{-2} \\
\\[-10pt]
=
T 
\!+\!
{\displaystyle \frac{4\hbar ^2 }{\mu  N r^2}}
\!-\!
{\displaystyle \frac{r^2 }{2\mu N r_0 ^4}} \!
\sum^{2}_{i \!=\! 1} \pi _i \pi _i 
\!-\!
{\displaystyle \frac{16\hbar ^2 r_0 ^4}{\mu N r^6}} \!
\sum^{2}_{i \!=\! 1} \phi _i \phi _i 
\!+\!
i\hbar {\displaystyle \frac{2}{\mu N r^2}} \!
\sum^{2}_{i \!=\! 1} \phi _i \pi _i  \\
\\[-10pt]
-
{\displaystyle \frac{\hbar ^2 r_0 ^{2}}{2\mu}} \!
\left( \!
2  
\!+\!
{\displaystyle \frac{166}{3N}} \!
\right) \!
r_0 ^{-2} r^{-2} .
\end{array}
\label{constC0}
\eeqa
Substituting
(\ref{sumphipi})
and
(\ref{sumpipi})
into
(\ref{constC0}),
$C_{0}(r^{2}) $
is expressed as\\[-16pt]
\beqa
\begin{array}{ll}
C_{0}(r^{2})  
\!=\!
T 
\!-\!
{\displaystyle \frac{r^2 }{2\mu N r_0 ^4}} \!
\left\{ \!{}^{^{^{^{^{^{}}}}}}\!
\hbar^{2} 4 r^{4}_{0} r^{-4}
\!-\! 
\hbar^{2} 20 r^{8}_{0} r^{-8} \!
\sum^{2}_{j \!=\! 1} \phi _j  \phi _j 
\right. \\
\\[-10pt]
\left.
+
\hbar^{2} 2 r^{4}_{0} r^{-4} \!
\sum_{n=1}^N \!\!
\left( \!
{\displaystyle 
x_n \frac{\partial}{\partial x_n} 
\!+\!
y_n \frac{\partial}{\partial y_n}
} \!
\right) \!
\!-\!
\hbar^{2} N r^{4}_{0} r^{-2} \!
\sum_{n=1}^N \!\!
\left( \!
{\displaystyle 
\frac{\partial^{2}}{\partial x_n^{2}} 
\!+\!
\frac{\partial^{2}}{\partial y_n^{2}}
} \!
\right) \!\! 
\right\} \\
\\[-10pt]
\!+\!
i\hbar {\displaystyle \frac{2}{\mu N r^2}} \!
\left\{ \!
- i \hbar
{\displaystyle \frac{1}{2}} \!
\sum_{n=1}^N \!\!
\left( \!
{\displaystyle 
x_n \frac{\partial}{\partial x_n} 
\!+\!
y_n \frac{\partial}{\partial y_n}
} \!
\right) \!
\!+\!
i \hbar 2 r^{4}_{0} r^{-4} \!
\sum^{2}_{j \!=\! 1} \phi _j  \phi _j \!
\right\} \\
\\[-8pt]
+
{\displaystyle \frac{\hbar ^2 r_0 ^{2}}{2\mu }}
{\displaystyle \frac{8}{N}}
r_0 ^{-2} r^{-2}
\!-\!
{\displaystyle \frac{\hbar ^2 r_0 ^{2}}{2\mu }}
{\displaystyle \frac{32 }{N}}
r_0 ^{2} r^{-6}
\sum^{2}_{i \!=\! 1} \phi _i \phi _i 
-
{\displaystyle \frac{\hbar ^2 r_0 ^{2}}{2\mu}} \!
\left( \!
2  
\!+\!
{\displaystyle \frac{166}{3N}} \!
\right) \!
r_0 ^{-2} r^{-2} ,
\end{array}
\label{C01}
\eeqa\\[-10pt]
which
is rearranged as\\[-20pt]
\beqa
\begin{array}{ll}
C_{0}(r^{2})  
=
T
\!+\!
{\displaystyle \frac{\hbar^{2}}{2\mu }} \!
\sum_{n=1}^N \!
\left( \!
{\displaystyle 
\frac{\partial^{2}}{\partial x_n^{2}} 
\!+\!
\frac{\partial^{2}}{\partial y_n^{2}}
} \!
\right) \\
\\[-10pt]   
-
{\displaystyle \frac{\hbar^{2} r^{2}_{0}}{2\mu }} \!
{\displaystyle \frac{2}{N}}
r^{-2}_{0} r^{-2} \!
\left\{ \!
\sum_{n=1}^N \!
\left( \!
{\displaystyle 
x_n \frac{\partial}{\partial x_n} 
\!+\!
y_n \frac{\partial}{\partial y_n}
} \!
\right) 
\!-\!
\sum_{n=1}^N \!
\left( \!
{\displaystyle 
x_n \frac{\partial}{\partial x_n} 
\!+\!
y_n \frac{\partial}{\partial y_n}
} \!\!
\right) \!
\right\} \\
\\[-10pt]
-
{\displaystyle \frac{\hbar^{2} r^{2}_{0}}{2\mu }} \!
{\displaystyle \frac{2}{N}}
r^{-2}_{0} r^{-2} \!
\left\{ \!
2 
\!-\! 
10
r^{4}_{0} r^{-4} \!
\sum^{2}_{j \!=\! 1} \! \phi _j  \phi _j \!
\right\}
\!-\!
{\displaystyle \frac{\hbar ^2 r_0 ^{2}}{2\mu }}
{\displaystyle \frac{8}{N}}
r^{2}_{0} r^{-6} \!
\sum^{2}_{j \!=\! 1} \phi _j  \phi _j \\
\\[-8pt]
+
{\displaystyle \frac{\hbar ^2 r_0 ^{2}}{2\mu }}
r_0 ^{-2} r^{-2} \!
\left\{ \!
{\displaystyle \frac{8}{N}}
\!-\!
{\displaystyle \frac{32 }{N}}
r_0 ^{4} r^{-4} \!
\sum^{2}_{i \!=\! 1} \phi _i \phi _i 
\!-\!
\left( \!\!
2  
\!+\!
{\displaystyle \frac{166}{3N}} \!
\right) \!
\right\} ,
\end{array}
\label{constC01}
\eeqa
in the R.H.S. of the previous equation
the terms in the first and second lines cancel each other.
Using again the approximate relation
$
\sum^{2}_{i \!=\! 1} \! \phi _i  \phi _i
\!\!\approx\!\!
{\displaystyle \frac{1}{4}}
r_0 ^{-4} r^{4}
$,
the constant term  
$C_{0}(r^{2}) $
is determined as\\[-22pt]
\beqa
C_{0}(r^{2})  
\!=\!
-
{\displaystyle \frac{\hbar ^2 r_0 ^{2}}{2\mu}} \!
\left( \!\!
2  
\!+\!
{\displaystyle \frac{169}{3N}} \!
\right) \!
 r_0 ^{-2} r^{-2} . 
\label{C0determined}
\eeqa\\[-10pt]
Thus, the final expression for the kinetic part  $T$ of the Hamiltonian
is given as follows:\\[-18pt]
\beqa
\!\!\!\!\!\!\!\!
\left.
\BA{lr}
T
\!=\!
T^{(0)} (\phi ; r^2 ) 
\!-\!
{\displaystyle \frac{4 \hbar ^2 }{\mu N r^2}} 
\!+\!
{\displaystyle \frac{r^2 }{2\mu N r_0 ^4}} \!
\sum^{2}_{i \!=\! 1} \pi _i \pi _i 
\!+\!
{\displaystyle \frac{16\hbar ^2 r_0 ^4}{\mu N r^6}} \!
\sum^{2}_{i \!=\! 1} \phi _i \phi _i 
\!-\!
i\hbar {\displaystyle \frac{2}{\mu N r^2}} \!
\sum^{2}_{i \!=\! 1} \phi _i \pi _i  , \\
\\[-12pt]
T^{(0)}(\phi ; r^2) 
\!=\!
C_{0}(r^{2}) 
+
{\displaystyle \frac{\hbar ^2 r_0 ^{2}}{2\mu}} 
2 \!
\left( \! 4 \!-\! {\displaystyle \frac{3}{N}} \! \right) \!
r_0 ^2 r^{-6} \! 
\sum^{2}_{i \!=\! 1} \phi _i  \phi _i \!
+
{\displaystyle \frac{\hbar ^2 r_0 ^{2}}{2\mu}}
{\displaystyle \frac{2728}{3N}} \!
r_0 ^6 r^{-10} \!
\left( \sum^{2}_{i \!=\! 1} \phi _i  \phi _i \right)^{2} \! ,  \\
\\[-12pt]
C_{0}(r^{2})  
\!=\!
-
{\displaystyle \frac{\hbar ^2 r_0 ^{2}}{2\mu}} \!
\left( \!\!
2  
\!+\!
{\displaystyle \frac{169}{3N}} \!
\right) \!
 r_0 ^{-2} r^{-2} .
\EA \!\!
\right\}
\label{Tdeterminedpiphi}
\eeqa\\[-10pt]
As mentioned in the end of the previous section,
we emphasize again that
the total kinetic energy $T$ including the term $T^{(0)}(\phi ; r^2)$
gives a strong anharmonicity in the vibrational regime.
Specifically,
the coefficients of $\phi_1^2$ or $\phi_2^2$ may be negative.
More explicitly,
the quadratic form involving $\phi_1\phi_2$, $\phi_1^2$ and $\phi_2^2$
will not be positive definite.

To get the above final expression
(\ref{Tdeterminedpiphi}),
many approximations were made. 
In particular,
the approximate mean-value
$\sum_{n=1}^N \!
{\displaystyle 
\langle
x_n \frac{\partial}{\partial x_n} 
\!+\!
y_n \frac{\partial}{\partial y_n}
\rangle
}
\!\!=\!\!
f(N)$
in
(\ref{sumphietameanvalue})
plays a crucial role to make the commutator
$\left[ \pi_i, T^{(0)} \! (\phi ;r^2) \right]$
to be dependent only on the variables
$\phi_1$ and $\phi_2$.
The verification is made through the procedure:
$\!\!
\left\{ \!
{\displaystyle \frac{1}{N}} f(N)
\!+\!
\left( \! 1 \!-\! {\displaystyle \frac{1}{N}} \! \right) \!
\right\} \!
\eta _i
\ket{\mbox{coll.subspace}}
\!=\!
0
\!\Rightarrow\!
f(N)
\!=\!
- N \!+\! 1
$.
This manner means that
such an approximation
does not act directly on the $\ket{\mbox{coll.subspace}}$
but the coefficient of the operator makes to vanish.
The other approximations are also made in the same way
as the the above way.

\newpage


\def\thesection{\arabic{section}}
\setcounter{equation}{0}
\renewcommand{\theequation}{\arabic{section}.
\arabic{equation}}     

\section{Discussion and concluding remarks}

~~~As for the present two-dimensional nuclei,
particularly,
we study the structure of the collective subspace
satisfying the subsidiary condition
(\ref{subsidiarycondition}).
This subsidiary condition
is important to investigate the structure of
the collective subspace.
We denote a wave function of
the collective subspace
$ \langle \mbox{coll.subspace} \ket \Psi $
as
${\Psi} [\cdot]$.
From now on,
we investigate various possibilities
satisfying the subsidiary condition.
First, as a trial, let us consider a certain wave function
belonging to a subspace of the Hilbert space.
We express it as
${\Phi} [\cdot]$.
On this wave function,
we assume an {\bf ansatz}:
$
{\Phi} [\cdot]
\!=\!
{\Phi} [r]
$.
Then on the ${\Phi} [r]$ we have\\[-16pt]
\beqa
\BA{c}
{\displaystyle
\frac{\partial}{\partial x_n} {\Phi} [r] 
\!=\! 
\frac{1}{N} \frac{x_n}{r} 
\frac{\partial}{\partial r} {\Phi} [r] ,~~
\frac{\partial}{\partial y_n} { \Phi} [r]
\!=\!
\frac{1}{N} \frac{y_n}{r} 
\frac{\partial}{\partial r} 
{\Phi} [r]
} ,~~
\left(
r
\!=\!
\sqrt{
{\displaystyle \frac{1}{N }} \!
\sum^{N}_{n \!=\! 1}
\left( 
x^{2}_n 
\!+\!
y^{2}_n  
\right)
}
\right) ,
\EA
\label{anzatz} 
\eeqa\\[-12pt]
Now we introduce an auxiliary operator $\widetilde{\bm L}$
which is expressed as\\[-18pt] 
\beqa
\widetilde{\bm L}
\!=\!
l \!-\! {\bm L}_{\mbox{coll}}
,~~~
{\bm L}_{\mbox{coll}} \!\equiv\! 2 ( \phi_1 \pi_2 
\!-\! \phi_2 \pi_1 )  .
\label{subsidiarycondition2}
\eeqa\\[-18pt] 
Substituting the explicit expression for the total angular momentum operator $l$
and the second relation of 
(\ref{newoldrelation})
into
(\ref{subsidiarycondition2}),
an action of the $\widetilde{\bm L}$ onto the ${\Phi} [r]$
is calculated as follows:\\[-16pt] 
\beqa
\!\!
\BA{lr}
\widetilde{\bm L} \! {\Phi} [r]
\!\!=\!\!
\left[ 
{\displaystyle \frac{ \hbar }{i}} \!
\sum_n \!\!
\left( \!\!
{\displaystyle 
x_n \frac{\partial}{\partial y_n}
\!\!-\!\!
y_n \frac{\partial}{\partial x_n} 
} \!\!
\right) 
\!-\!
2 r_0 ^2 r^{-2} \!
\left\{
\phi_1 \!
\left(  
{\eta}_{2}
\!+\!
i\hbar 
2 r_0 ^2 r^{-2} \! {\phi}_{2} 
\right) 
\!-\!
\phi_2 \!
\left( 
{\eta}_{1}
\!+\!
i\hbar 
2 r_0 ^2 r^{-2} \! {\phi}_{1}
\right) 
\right\}
{}^{^{^{^{^{^{}}}}}}\!\!\!
\right] \!\!
{\Phi} [r] \\
\\[-12pt]
\!=\!
{\displaystyle \frac{ \hbar }{i}} \!
\left[ \!
{\displaystyle \frac{1}{N}}
{\displaystyle \frac{1}{r}} \!
\sum_n \!\!
\left( \!
{\displaystyle 
x_n y_n
\!-\!
y_n x_n 
} \!
\right) 
\!-\!
2 r_0 ^2 r^{-2} \!\!
\left\{ \!\!
\phi_1 \!\!
\sum_n \!\!
\left( \!\!
{\displaystyle 
x_n \frac{\partial}{\partial y_n}
\!+\!
y_n \frac{\partial}{\partial x_n} 
} \!\!
\right) \!
\!-\!
\phi_2 \!\!
\sum_n \!\!
\left( \!\!
{\displaystyle 
x_n \frac{\partial}{\partial y_n}
\!-\!
y_n \frac{\partial}{\partial x_n} 
} \!\!
\right) \!\!
\right\} \!
{}^{^{^{^{^{^{}}}}}}\!\!\!
\right] \!\!
{\Phi} [r] \\
\\[-12pt]
\!=\!
-
{\displaystyle \frac{ \hbar }{i}}
2 r_0 ^2 r^{-2} \!
{\displaystyle \frac{1}{N}}
{\displaystyle \frac{1}{r}} \!
\left[
\phi_1 \!
\sum_n \!
\left(  
x_n y_n
\!+\!
y_n x_n 
\right) \!
{\displaystyle \frac{\partial}{\partial r}} 
\!-\!
\phi_2 \!
\sum_n \!
\left( 
x^{2}_n 
\!-\!
y^{2}_n  
\right) \!
{\displaystyle \frac{\partial}{\partial r}} 
\right] \!\!
{\Phi} [r] \\
\\[-12pt]
\!=\!
-
{\displaystyle \frac{ \hbar }{i}}
4 r_0 ^4 r^{-3} \!
\left[
\phi_1 
\!\cdot\!
{\displaystyle \frac{1}{N r_0 ^2}} \!
\sum_n \!  
x_n y_n
{\displaystyle \frac{\partial}{\partial r}} 
\!-\!
\phi_2 
\!\cdot\!
{\displaystyle \frac{1}{N r_0 ^2}} \!
\sum_n \!
{\displaystyle \frac{1}{2}}
\left( 
x^{2}_n 
\!-\!
y^{2}_n  
\right) \!
{\displaystyle \frac{\partial}{\partial r}} 
\right] \!\!
{\Phi} [r] \\
\\[-16pt]
\!=\!
4 i\hbar
r_0 ^4 r^{-3} \!
\left[
\phi_1 
\phi_2
\!-\!
\phi_2 
\phi_1 
\right] \!
{\displaystyle \frac{\partial}{\partial r}} 
{\Phi} [r] 
=
0 ,
\label{actPhi}
\EA
\eeqa\\[-14pt]
where we have used the relations
(\ref{anzatz})
and the collective coordinates
(\ref{Collectivecoordinates})
and cojugate momenta
(\ref{Conjugatemomenta2}).
The {\bf ansatz} 
$
{\Phi} [\cdot]
\!=\!
{\Phi} [r]
$
is one of the possible solutions for the 
wave function of the collective subspace
from a mathematical point a view.
The variable $r$, however, is not a collective operator
but an intrinsic operator.
Second,
on the $\ket{\mbox{coll.subspace}}$,
the $exact$ canonical momentum $\pi_i$
is represented as
${\displaystyle \frac{ \hbar }{i}\frac{\partial}{\partial \phi_i}}$.
Then, 
the subsidiary condition
$\widetilde{\bm L}{\Psi} [\cdot]
\!=\!
0$
on ${\Psi} [\phi_1, \phi_2]$ is computed as\\[-16pt]
\beqa
\!\!
\BA{lr}
\widetilde{\bm L}{\Psi} [\phi_1, \phi_2]
\!=\!
\left[
{\displaystyle \frac{ \hbar }{i}} 
\sum_n \!
\left( \!
{\displaystyle 
x_n \frac{\partial}{\partial y_n}
\!-\!
y_n \frac{\partial}{\partial x_n} 
} \!
\right) 
\!-\!
2 {\displaystyle \frac{ \hbar }{i}} \!
\left(  \!
\phi_1
{\displaystyle \frac{\partial}{\partial \phi_2}}
\!-\!
\phi_2
{\displaystyle \frac{\partial}{\partial \phi_1}} \!
\right) 
\right] \!
{\Psi} [\phi_1, \phi_2] \\
\\[-8pt]
\!=\!
{\displaystyle \frac{ \hbar }{i}} \!
\left[ \!
\sum_{n} \!\!
\left( \!\!
{\displaystyle 
x_n \frac{\partial \phi_1}{\partial y_n}
\frac{\partial}{\partial \phi_1}
\!\!+\!\!
x_n \frac{\partial \phi_2}{\partial y_n}
\frac{\partial}{\partial \phi_2}
\!-\!
y_n \frac{\partial \phi_1}{\partial x_n}
\frac{\partial}{\partial \phi_1}
\!-\!
y_n \frac{\partial \phi_2}{\partial x_n}
\frac{\partial}{\partial \phi_2} \!
} \!
\right) \!
\!\!-\!\!
2 \!
\left(  \!\!
\phi_1
{\displaystyle \frac{\partial}{\partial \phi_2}}
\!-\!
\phi_2
{\displaystyle \frac{\partial}{\partial \phi_1}} \!
\right) \! 
\right] \!\!
{\Psi} [\phi_1, \phi_2] \\
\\[-8pt]
\!=\!
{\displaystyle \frac{ \hbar }{i}} \!
\left[ \!
{\displaystyle \frac{1}{N r_0 ^2}} \!
\sum_{n} \!\!
\left( \!
{\displaystyle 
-x_n y_n
\frac{\partial}{\partial \phi_1}
\!+\!
x_n ^2
\frac{\partial}{\partial \phi_2}
\!-\!
y_n x_n
\frac{\partial}{\partial \phi_1}
\!-\!
y_n ^2
\frac{\partial}{\partial \phi_2} \!
} \!
\right) \!
\!-\!
2 \!\!
\left(  \!\!
\phi_1
{\displaystyle \frac{\partial}{\partial \phi_2}}
\!-\!
\phi_2
{\displaystyle \frac{\partial}{\partial \phi_1}} \!
\right) \! 
\right] \!\!
{\Psi} [\phi_1, \phi_2] 
\!=\!
0,
\EA
\label{actPsi}
\eeqa\\[-10pt]
which
shows
${\Psi} [\cdot]
\!=\!
{\Psi} [\phi_1, \phi_2]
$
is a solution to the problem.
Finally,
on a space expressed by the microscopic degrees of freedom
$\{(x_i,y_i)| i \!=\! 1,\! \cdots\!, N\!\}$,
the subsidiary condition is written as \\[-16pt]
\beqa
\!\!\!\!
\BA{ll}
\widetilde{\bm L}
\Psi[x_1,y_1,...,x_n,y_n]
\!=\!
\left[
l - 2 r_0^2 r^{-2} \!
\left\{
\phi_1 \eta_2
\!-\!
\phi_2 \eta_1
\right\}
\right]
\Psi[x_1,y_1,...,x_n,y_n] \\
\\[-10pt]
\!=\!\!
{\displaystyle \frac{ \hbar }{i}} \!\!
\sum_n \!\!
\left[ \!
\left( \!\!
x_n 
{\displaystyle
\frac{\partial}{\partial y_n}
} \!
\!-\!
y_n
{\displaystyle 
 \frac{\partial}{\partial x_n}
} \!\!
\right)  
\!\!-\!\!
2  r_{\!0}^2 \! r^{\!-2} \!\! 
\left\{  \!\!
\phi_1 \!\!
\left( \!\!
{\displaystyle 
x_n \! \frac{\partial}{\partial x_n}
\!\!+\!\!
y_n \! \frac{\partial}{\partial y_n} 
} \!\!
\right) 
\!\!-\!\!
\phi_2 \!\!
\left( \!\!
{\displaystyle 
x_n \! \frac{\partial}{\partial x_n}
\!\!-\!\!
y_n \! \frac{\partial}{\partial y_n} 
} \!\!
\right)  \!\!
\right\} \!
\right] \!\!
\Psi[x_1,y_1,...,x_n,y_n] \\
\\[-10pt]
\!=\!\!
{\displaystyle \frac{ \hbar }{i}} \!\!
\sum_n \!\!
\left[
\left\{
- y_n
\!\!-\!\!
2  r_{\!0}^2 \! r^{\!-2} \!
\left( \!
\phi_1
\!-\!
\phi_2 \!
\right) 
x_n
\right\} \!\!
{\displaystyle 
 \frac{\partial}
{\partial x_n}
\!+\!\!
\left\{
x_n  
\!\!-\!\!
2  r_{\!0}^2 \! r^{\!-2} \! 
\left( \!
\phi_1
\!+\!
\phi_2 \!
\right) 
y_n
\right\} \!\!
\frac{\partial}{\partial y_n} 
} \!
\right] \!\!
\Psi[x_1,y_1,...,x_n,y_n]
\!=\!
0.
\EA
\label{diffeq}
\eeqa
Thus, we can obtain a differential equation in the microscopic variables
$\{(x_i,y_i)|i \!=\! 1\cdots,N\}$ only.
This is just the differential equation we want
when trying to construct the collective space
from microscopic considerations only.
As is clear from the structure of the equation
(\ref{diffeq}),
the $\Phi[r]$ and ${\Psi} [\phi_1, \phi_2]$
are not general solutions of the differential equation.
However, there may be other solutions than them
to satisfy the subsidiary condition.

In this paper, 
the $exact$ canonical momenta
$\pi_i$
to
$\phi_i$
is derived by modifying the approximate momenta
$\eta_i$.
The commutativity of momenta
$\pi_i$ and $\pi_j$ is shown
under the subsidiary condition.
Using the $exact$ canonical variables
$\phi_i$
and
$\pi_i$,
we found the collective Hamiltonian
including the so-called phonon-phonon interaction
under the replacement of operator $r^2$,
which is not a collective variable,
by an average value
$
\langle r^2 \rangle
\!\!=\!\!
r^{2}_{0}$.
In the simple two-dimensional nuclei,
discussions of the couplings between the individual particle motion
and the collective motion will be possible
if we investigate in detail the collective subspace
relating to the individual particle motion through the variable $r^2$.
This work will be presented elsewhere.

\vspace{0.6cm}


\centerline{\bf Acknowledgements}
\vspace{0.2cm}
S. N. would like to
express his sincere thanks to 
Professor Constan\c{c}a Provid\^{e}ncia for kind and
warm hospitality extended to
him at the Centro de F\'\i sica,
Universidade de Coimbra, Portugal.
This work was supported by FCT (Portugal) under the project
CERN/FP/83505/2008.

\newpage

\leftline{\large{\bf Appendix}} 
\appendix
\def\thesection{\Alph{section}}
\setcounter{equation}{0}  
\renewcommand{\theequation}{\Alph{section}.\arabic{equation}}
\section{Derivation of (\ref{relationeta})}

~~~~We here derive the approximate relation
(\ref{relationeta})
for the case of 
$\eta_{1}$.\\[-14pt]
\beqa
\BA{lr}
\sum_{n=1}^N \!
\left( \!
{\displaystyle 
x_n \frac{\partial}{\partial x_n} 
\!+\!
y_n \frac{\partial}{\partial y_n}
} \!
\right) \!
\eta _1  
=
- i\hbar 
\sum_{n, n^{\prime}=1}^N \!
\left( \!
{\displaystyle 
x_n \frac{\partial}{\partial x_n} 
\!+\!
y_n \frac{\partial}{\partial y_n}
} \!
\right) \!
\left( \!
{\displaystyle 
x_{n^{\prime}} \frac{\partial}{\partial x_{n^{\prime}}} 
\!-\!
y_{n^{\prime}} \frac{\partial}{\partial y_{n^{\prime}}}
} \!
\right) \\
\\[-6pt]
= - i\hbar 
\sum_{n, n^{\prime}=1}^N \!
\left\{ \!
x_n
\left( \!
\left[ {\displaystyle \frac{\partial}{\partial x_n}}, x_{n^{\prime}} \right]
+
x_{n^{\prime}} {\displaystyle \frac{\partial}{\partial x_n}}
\right) \!
{\displaystyle \frac{\partial}{\partial x_{n^{\prime}}}}
-
{\displaystyle x_n y_{n^{\prime}} \frac{\partial}{\partial x_n}} \!
{\displaystyle \frac{\partial}{\partial y_{n^{\prime}}}}
\right. \\
\\[-6pt]
\left.
~~~~~~~~~~~~~~~~~~~~~~~~~~~~~~~~~~~~~~~~\! 
+
{\displaystyle y_n x_{n^{\prime}} \frac{\partial}{\partial y_n}} \!
{\displaystyle \frac{\partial}{\partial x_{n^{\prime}}}}
-
y_n
\left( \!
\left[ {\displaystyle \frac{\partial}{\partial y_n}}, y_{n^{\prime}} \right]
+
y_{n^{\prime}} {\displaystyle \frac{\partial}{\partial y_n}}
\right) \! 
{\displaystyle \frac{\partial}{\partial y_{n^{\prime}}}}
\right\} \\
\\[-6pt]
\approx 
- i\hbar 
\sum_{n, n^{\prime}=1}^N \!
\left\{ \!
\langle
x_n
{\displaystyle \frac{\partial}{\partial x_n}}
\rangle
x_{n^{\prime}} 
{\displaystyle \frac{\partial}{\partial x_{n^{\prime}}}}
-
\langle
x_n
{\displaystyle \frac{\partial}{\partial x_n}}
\rangle
{\displaystyle y_{n^{\prime}} \frac{\partial}{\partial y_{n^{\prime}}}}
\right. \\
\\[-6pt]
\left.
~~~~~~~~~~~~~~~~~~~~~~~~~~~~~~~~~~~~~~~~~~\! 
+
\langle
y_n
{\displaystyle \frac{\partial}{\partial y_n}}
\rangle
{\displaystyle x_{n^{\prime}} \frac{\partial}{\partial x_{n^{\prime}}}}
-
\langle
y_n
{\displaystyle \frac{\partial}{\partial y_n}}
\rangle
y_{n^{\prime}} {\displaystyle \frac{\partial}{\partial y_{n^{\prime}}}}
\right\} \\
\\[-6pt]
~~~~ 
- i\hbar
\left( \! 
\sum_{n=1}^N \!
x^{2}_n
\sum_{n^{\prime}=1}^N \!
{\displaystyle \frac{\partial}{\partial x^{2}_{n^{\prime}}}}
-
\sum_{n=1}^N \!
y^{2}_n
\sum_{n^{\prime}=1}^N \!
{\displaystyle \frac{\partial}{\partial y^{2}_{n^{\prime}}}} \!
\right)
\!+\!
{\displaystyle \frac{\hbar}{i}}
\sum_{n=1}^N \!
\left( \!
{\displaystyle 
x_n \frac{\partial}{\partial x_n} 
\!-\!
y_n \frac{\partial}{\partial y_n}
} \!
\right) \\
\\[-6pt]
= 
- i\hbar
\left\{ \!
{\displaystyle \frac{1}{2}} \!
\sum_{n=1}^N \!
\left( \! x^{2}_n \!-\! y^{2}_n \! \right) \!
\sum_{n^{\prime}=1}^N \!
\left( \!
{\displaystyle \frac{\partial}{\partial x^{2}_{n^{\prime}}}}
\!+\!
{\displaystyle \frac{\partial}{\partial y^{2}_{n^{\prime}}}} \!
\right)
\!+\!
{\displaystyle \frac{1}{2}} \!
\sum_{n=1}^N \!
\left( \! x^{2}_n \!+\! y^{2}_n \! \right) \!
\sum_{n^{\prime}=1}^N \!
\left( \!
{\displaystyle \frac{\partial}{\partial x^{2}_{n^{\prime}}}}
\!-\!
{\displaystyle \frac{\partial}{\partial y^{2}_{n^{\prime}}}} \!
\right) \!
\right\} \\
\\[-6pt]
~~
+ \eta_1
\!+\!
\sum_{n=1}^N \!
\langle
x_n
{\displaystyle \frac{\partial}{\partial x_n}}
\!+\!
y_n
{\displaystyle \frac{\partial}{\partial y_n}}
\rangle
{\displaystyle \frac{\hbar}{i}}
\sum_{n^{\prime}=1}^N \!
\left( \!
{\displaystyle 
x_n^{\prime} \frac{\partial}{\partial x_n^{\prime}} 
\!-\!
y_n^{\prime} \frac{\partial}{\partial y_n^{\prime}}
} \!
\right) \! ,
\EA
\eeqa\\[-8pt]
in which
$
\langle
x_n
{\displaystyle \frac{\partial}{\partial x_n}}
\!+\!
y_n
{\displaystyle \frac{\partial}{\partial y_n}}
\rangle
$
stands for the mean-value of the operator
$\!
\left( \!
x_n
{\displaystyle \frac{\partial}{\partial x_n}}
\!+\!
y_n
{\displaystyle \frac{\partial}{\partial y_n}} \!
\right)
$
on the collective subspace
$\ket{\mbox{coll.subspace}}$.
\beqa
\BA{lr}
\sum_{n=1}^N \!
\left( \!
{\displaystyle 
x_n \frac{\partial}{\partial x_n} 
\!+\!
y_n \frac{\partial}{\partial y_n}
} \!
\right) \!
\eta _1  
=
\eta_1
\!+\!
\sum_{n=1}^N \!
\langle
x_n
{\displaystyle \frac{\partial}{\partial x_n}}
\!+\!
y_n
{\displaystyle \frac{\partial}{\partial y_n}}
\rangle
\eta_1 \\
\\[-6pt]
~  
- i\hbar
N r^{2}_{0}
{\displaystyle \frac{1}{2}} \!
\phi_1 \!
\sum_{n=1}^N \!
\left( \!
{\displaystyle \frac{\partial}{\partial x^{2}_{n}}}
\!+\!
{\displaystyle \frac{\partial}{\partial y^{2}_{n}}} \!
\right)
\!-\!
i\hbar
{\displaystyle \frac{1}{2}} \!
N r^{2}_{0} \!
\sum_{n=1}^N \!
\left( \!
{\displaystyle \frac{\partial}{\partial x^{2}_{n}}}
\!-\!
{\displaystyle \frac{\partial}{\partial y^{2}_{n}}} \!
\right) \\
\\[-6pt]
=
\eta_1
\!+\!
f(N)
\eta_1
\!+\!
{\displaystyle \frac{\mu}{\hbar^{2}}}
{\displaystyle \frac{1}{2}} \!
N r^{2} \!
\left[ \eta_1, T \right]
- i\hbar
N r^{2}_{0}
\phi_1 \!
\sum_{n=1}^N \!
\left( \!
{\displaystyle \frac{\partial}{\partial x^{2}_{n}}}
\!+\!
{\displaystyle \frac{\partial}{\partial y^{2}_{n}}} \!
\right) \! ,
\EA
\label{eta1relation1}
\eeqa
where we have used the first equation of
(\ref{Collectivecoordinates}),$\!$ (\ref{Conjugatemomenta1}), 
the first equation of$\!$
(\ref{Conjugatemomenta2})
and$\!$
(\ref{r**2andAngularmomentum}).
Then we have the relation
\beqa
\BA{cc}
\sum_{n=1}^N \!
\left( \!
{\displaystyle 
x_n \frac{\partial}{\partial x_n} 
\!+\!
y_n \frac{\partial}{\partial y_n}
} \!
\right) \!
\eta _1
=
\eta_1
\!+\!
f(N)
\eta_1
\!+\!
{\displaystyle \frac{\mu}{\hbar^{2}}}
{\displaystyle \frac{1}{2}} \!
N r^{2} \!
\left[ \eta_1, T \right] \\
\\[-6pt]
- i\hbar
N r^{2}_{0}
\phi_1 \!
\sum_{n=1}^N \!
\left( \!
{\displaystyle \frac{\partial}{\partial x^{2}_{n}}}
\!+\!
{\displaystyle \frac{\partial}{\partial y^{2}_{n}}} \!
\right) \! .
\EA
\label{eta1relation2}
\eeqa
In the above equations
(\ref{eta1relation1})
and
(\ref{eta1relation2}),
we have used the second relation of
(\ref{sumphietameanvalue})
for the unknown function $f(N)$.

\newpage

Next we derive the approximate relation
(\ref{relationeta})
for the case of 
$\eta_{2}$.\\[-14pt]
\beqa
\BA{lr}
\sum_{n=1}^N \!
\left( \!
{\displaystyle 
x_n \frac{\partial}{\partial x_n} 
\!+\!
y_n \frac{\partial}{\partial y_n}
} \!
\right) \!
\eta _2  
=
- i\hbar 
\sum_{n, n^{\prime}=1}^N \!
\left( \!
{\displaystyle 
x_n \frac{\partial}{\partial x_n} 
\!+\!
y_n \frac{\partial}{\partial y_n}
} \!
\right) \!
\left( \!
{\displaystyle 
x_{n^{\prime}} \frac{\partial}{\partial x_{n^{\prime}}} 
\!+\!
y_{n^{\prime}} \frac{\partial}{\partial y_{n^{\prime}}}
} \!
\right) \\
\\[-6pt]
= - i\hbar 
\sum_{n, n^{\prime}=1}^N \!
\left\{ \!
x_n
\left( \!
\left[ {\displaystyle \frac{\partial}{\partial x_n}}, x_{n^{\prime}} \right]
+
x_{n^{\prime}} {\displaystyle \frac{\partial}{\partial x_n}}
\right) \!
{\displaystyle \frac{\partial}{\partial x_{n^{\prime}}}}
+
{\displaystyle x_n y_{n^{\prime}} \frac{\partial}{\partial x_n}} \!
{\displaystyle \frac{\partial}{\partial y_{n^{\prime}}}}
\right. \\
\\[-6pt]
\left.
~~~~~~~~~~~~~~~~~~~~~~~~~~~~~~~~~~~~~~~~\! 
+
{\displaystyle y_n x_{n^{\prime}} \frac{\partial}{\partial y_n}} \!
{\displaystyle \frac{\partial}{\partial x_{n^{\prime}}}}
+
y_n
\left( \!
\left[ {\displaystyle \frac{\partial}{\partial y_n}}, y_{n^{\prime}} \right]
+
y_{n^{\prime}} {\displaystyle \frac{\partial}{\partial y_n}}
\right) \! 
{\displaystyle \frac{\partial}{\partial y_{n^{\prime}}}}
\right\} \\
\\[-6pt]
\approx 
- i\hbar 
\sum_{n, n^{\prime}=1}^N \!
\left\{ \!
\langle
x_n
{\displaystyle \frac{\partial}{\partial x_n}}
\rangle
x_{n^{\prime}} 
{\displaystyle \frac{\partial}{\partial x_{n^{\prime}}}}
+
\langle
x_n
{\displaystyle \frac{\partial}{\partial x_n}}
\rangle
{\displaystyle y_{n^{\prime}} \frac{\partial}{\partial y_{n^{\prime}}}}
\right. \\
\\[-6pt]
\left.
~~~~~~~~~~~~~~~~~~~~~~~~~~~~~~~~~~~~~~~~~~\! 
+
\langle
y_n
{\displaystyle \frac{\partial}{\partial y_n}}
\rangle
{\displaystyle x_{n^{\prime}} \frac{\partial}{\partial x_{n^{\prime}}}}
+
\langle
y_n
{\displaystyle \frac{\partial}{\partial y_n}}
\rangle
y_{n^{\prime}} {\displaystyle \frac{\partial}{\partial y_{n^{\prime}}}}
\right\} \\
\\[-6pt]
~~ 
-\! i\hbar
\left\{ \! 
\sum_{n=1}^N \!
x_n y_n \!
\sum_{n^{\prime}=1}^N \!
\left( \!
{\displaystyle \frac{\partial}{\partial x^{2}_{n^{\prime}}}}
\!+\!
{\displaystyle \frac{\partial}{\partial y^{2}_{n^{\prime}}}} \!
\right)
+
\sum_{n=1}^N \!
\left( \! x^{2}_n \!+\! y^{2}_n \! \right) \!
\sum_{n^{\prime}=1}^N \!
{\displaystyle \frac{\partial}{\partial x_{n^{\prime}}}} \!
{\displaystyle \frac{\partial}{\partial y_{n^{\prime}}}} \!
\right\} \\
\\[-6pt]
~~~
\!+
{\displaystyle \frac{\hbar}{i}}
\sum_{n=1}^N \!
\left( \!
{\displaystyle 
x_n \frac{\partial}{\partial x_n} 
\!+\!
y_n \frac{\partial}{\partial y_n}
} \!
\right) \\
\\[-6pt]
= 
- i\hbar
\left\{ \!
\sum_{n=1}^N \!
x_n y_n \!
\sum_{n^{\prime}=1}^N \!
\left( \!
{\displaystyle \frac{\partial}{\partial x^{2}_{n^{\prime}}}}
\!+\!
{\displaystyle \frac{\partial}{\partial y^{2}_{n^{\prime}}}} \!
\right)
\!+\!
{\displaystyle \frac{1}{2}} \!
\sum_{n=1}^N \!
\left( \! x^{2}_n \!+\! y^{2}_n \! \right) \!
\sum_{n^{\prime}=1}^N \!
\left( \!
{\displaystyle \frac{\partial}{\partial x_{n^{\prime}}}} \!
{\displaystyle \frac{\partial}{\partial y_{n^{\prime}}}}
\!+\!
{\displaystyle \frac{\partial}{\partial y_{n^{\prime}}}} \!
{\displaystyle \frac{\partial}{\partial x_{n^{\prime}}}} \!
\right) \!\!
\right\} \\
\\[-6pt]
~~
+ \eta_2
\!+\!
\sum_{n=1}^N \!
\langle
x_n
{\displaystyle \frac{\partial}{\partial x_n}}
\!+\!
y_n
{\displaystyle \frac{\partial}{\partial y_n}}
\rangle
{\displaystyle \frac{\hbar}{i}}
\sum_{n^{\prime}=1}^N \!
\left( \!
{\displaystyle 
x_n^{\prime} \frac{\partial}{\partial x_n^{\prime}} 
\!+\!
y_n^{\prime} \frac{\partial}{\partial y_n^{\prime}}
} \!
\right) \! ,
\EA
\eeqa\\[-8pt]
in which
$
\langle
x_n
{\displaystyle \frac{\partial}{\partial x_n}}
\!+\!
y_n
{\displaystyle \frac{\partial}{\partial y_n}}
\rangle
$
also stands for the mean-value of the operator
$\!
\left( \!
x_n
{\displaystyle \frac{\partial}{\partial x_n}}
\!+\!
y_n
{\displaystyle \frac{\partial}{\partial y_n}} \!
\right)
$
on the collective subspace
$\ket{\mbox{coll.subspace}}$.
Then we reach the final goal
\beqa
\BA{lr}
\sum_{n=1}^N \!
\left( \!
{\displaystyle 
x_n \frac{\partial}{\partial x_n} 
\!+\!
y_n \frac{\partial}{\partial y_n}
} \!
\right) \!
\eta _2  
=
\eta_2
\!+\!
\sum_{n=1}^N \!
\langle
x_n
{\displaystyle \frac{\partial}{\partial x_n}}
\!+\!
y_n
{\displaystyle \frac{\partial}{\partial y_n}}
\rangle
\eta_2 \\
\\[-6pt]
~  
- i\hbar
N r^{2}_{0}
\phi_2 \!
\sum_{n=1}^N \!
\left( \!
{\displaystyle \frac{\partial}{\partial x^{2}_{n}}}
\!+\!
{\displaystyle \frac{\partial}{\partial y^{2}_{n}}} \!
\right)
\!-\!
i\hbar
{\displaystyle \frac{1}{2}} \!
N r^{2} \!
\sum_{n=1}^N \!
\left( \!
{\displaystyle \frac{\partial}{\partial x_n}}
{\displaystyle \frac{\partial}{\partial y_n}}
\!+\!
{\displaystyle \frac{\partial}{\partial y_n}}
{\displaystyle \frac{\partial}{\partial x_n}} \!
\right) \\
\\[-6pt]
=
\eta_2
\!+\!
f(N)
\eta_2
\!+\!
{\displaystyle \frac{\mu}{\hbar^{2}}}
{\displaystyle \frac{1}{2}}
N r^{2} \!
\left[ \eta_2, T \right]
- i\hbar
N r^{2}_{0}
\phi_2 \!
\sum_{n=1}^N \!
\left( \!
{\displaystyle \frac{\partial}{\partial x^{2}_{n}}}
\!+\!
{\displaystyle \frac{\partial}{\partial y^{2}_{n}}} \!
\right) \! ,
\EA
\label{eta2relation1}
\eeqa
where we have used the second equation of
(\ref{Collectivecoordinates}), (\ref{Conjugatemomenta1}), 
the second equation of
(\ref{Conjugatemomenta2})
and
(\ref{r**2andAngularmomentum}).
Finally, we obtain the relation
\beqa
\BA{cc}
\sum_{n=1}^N \!
\left( \!
{\displaystyle 
x_n \frac{\partial}{\partial x_n} 
\!+\!
y_n \frac{\partial}{\partial y_n}
} \!
\right) \!
\eta _2
=
\eta_2
\!+\!
f(N)
\eta_2
\!+\!
{\displaystyle \frac{\mu}{\hbar^{2}}}
{\displaystyle \frac{1}{2}} \!
N r^{2} \!
\left[ \eta_2, T \right] \\
\\[-6pt]
- i\hbar
N r^{2}_{0}
\phi_2 \!
\sum_{n=1}^N \!
\left( \!
{\displaystyle \frac{\partial}{\partial x^{2}_{n}}}
\!+\!
{\displaystyle \frac{\partial}{\partial y^{2}_{n}}} \!
\right) \! .
\EA
\label{eta2relation2}
\eeqa
In the above equations$\!$
(\ref{eta2relation1})$\!$
and$\!$
(\ref{eta2relation2}),
we also have used the second relation of
(\ref{sumphietameanvalue})
for the $f(N)$.
Both the equations
(\ref{eta1relation2}) and (\ref{eta2relation2})
are combined into a single equation.
Thus we can derive the approximate relation
(\ref{relationeta}).

\end{document}